\documentstyle[preprint,aps,eqsecnum,prd,epsf,rotate]{revtex}

\begin{document}
\draft
\tighten

\title{
\vskip-3cm{\baselineskip14pt}
\centerline{\normalsize\hskip12.5cm hep-ph/9507413}
\centerline{\normalsize\hskip12.5cm TUM--HEP--223/95}
\centerline{\normalsize\hskip12.5cm Revised version:}
\centerline{\normalsize\hskip12.5cm To appear in PRD}
\vskip1.5cm
Large uncertainties in the cross section\\ 
of elastic $W_L^+W_L^-$ scattering 
}
\author{
Kurt Riesselmann\thanks{Electronic address: kurtr@physik.tu-muenchen.de}\,
}

\address{
Physik-Department T30, Technische Universit\"at M\"unchen,\\
James-Franck-Stra\ss e, 85747 Garching b.\ M\"unchen, Germany}

\maketitle

\begin{abstract}
  The Standard Model amplitudes for $2\rightarrow 2$ scattering processes
  involving longitudinally polarized gauge bosons $\left(W_L^\pm,\,Z_L\right)$
  and the Higgs boson are analyzed up to two loops. Assuming $M_H\gg M_W$, the
  trilinear Higgs coupling, $\lambda v$, is dominant for energies of
  $\sqrt{s}<$ 1.5 -- 2 $M_H$. For larger values of $\sqrt{s}$, the quartic
  coupling, $\lambda$, becomes dominant, allowing for a simpler calculation of
  higher-order corrections.  The resulting high-energy amplitudes display a
  large logarithmic dependence on $\sqrt{s}$ which can be resummed using
  renormalization group techniques.  For $M_H<350$ GeV, a next-to-leading-log
  calculation is sufficient.  For $350<M_H<450$ GeV, a
  next-to-next-to-leading-log calculation is necessary to include large
  two-loop corrections.  For a Higgs mass larger than ${\rm O}(450\, {\rm 
  GeV})$ and $\sqrt{s}>2M_H$, the perturbative results are not reliable.
  Choosing the $\overline{{\rm MS}}$ renormalization scheme instead of the OMS
  scheme, the coefficients of the perturbative series increase in magnitude,
  making the breakdown of perturbation theory even more apparent.  In
  conclusion, the perturbative cross sections presented here show very large
  uncertainties if $M_H\gtrsim450$ GeV and $\sqrt{s}\gtrsim 2\,M_H$, reducing
  the sensitivity to contributions from new physics significantly.
\end{abstract}

\narrowtext

\section{INTRODUCTION}
 
In the Standard Model, the weak gauge bosons $W^\pm$ and $Z$ acquire a mass by
means of the Higgs mechanism~\cite{higgs}.  Though the masses $M_W$ and $M_Z$
are experimentally well-known, the Higgs particle itself has not yet been
observed. This leaves the Higgs mass to be the last undetermined parameter of
the Standard Model.  Knowing the Higgs mass, $M_H$, the Higgs quartic coupling,
$\lambda=G_F\,M_H^2/\sqrt{2}$, and the Higgs trilinear coupling, $\lambda v
=G_F^{1/2}\,M_H^2/8^{1/4}$, are fixed.  All three quantities are intimately
connected to the scattering of longitudinally polarized gauge bosons $W_L^\pm$
and $Z_L$: The Higgs mass corresponds to the pole of the cross section, and the
Higgs couplings present the dominant contributions to the cross section if
$M_H\gg M_W$.  To test the Standard Model Higgs sector, it is therefore
important to analyze the cross sections of elastic $2\rightarrow 2$ processes
involving longitudinally polarized gauge bosons and the Higgs boson in more
detail.  Only if we understand the Standard Model scattering processes we are
able to test the Higgs couplings phenomenologically. Large uncertainties in the
Standard Model cross sections decrease the experimental sensitivity to new
physics significantly.

Assuming $\sqrt{s},M_H\gg M_W$, we discuss the amplitudes for $2\rightarrow 2$
scattering processes involving longitudinally polarized gauge bosons
$\left(W_L^\pm,\,Z_L\right)$ and the Higgs boson.  We briefly review the
one-loop results and examine the validity of the high-energy approximation.  We
present the two-loop high-energy amplitudes, and find both the logarithmic and
the non-logarithmic corrections to be important if $M_H$ is large.  Using
renormalization group techniques, the logarithms are resummed including the
complete set of next-to-next-to-leading logarithms (NNLL).  This gives a
perturbative series in the running coupling. We find the perturbative character
of this series to break down if $M_H\gtrsim 450$ GeV and $\sqrt{s}\gtrsim
2M_H$.

\section{FRAMEWORK}

For electroweak processes in which both $\sqrt{s}\gg M_W$ and $M_H\gg M_W$, the
electroweak interactions are dominated by the coupling of the longitudinal
components of the vector bosons, $W_L^\pm,Z_L$, to each other and to other
particles (leptons, quarks, or Higgs particle).  This is known as the
equivalence theorem (EQT)~\cite{eqt,lqt}.  In this limit, the dominant coupling
constants are the Higgs quartic coupling $\lambda$, the Higgs trilinear
coupling $\lambda v$, and the Yukawa couplings of the heavy fermions.  The
quantity $v$ is the vacuum expectation value of the Higgs sector,
$v=2^{-1/4}G_F^{-1/2}=246$ GeV.  Choosing the appropriate renormalization
scheme~\cite{eqtren}, the subdominant electroweak gauge couplings can be
neglected. Setting $g_1=g_2=0$, the longitudinal components of the electroweak
gauge bosons can be identified as the three massless Goldstone bosons,
$w^+,w^-$, and $z$ of the Higgs sector.  In this limit, all interactions are
determined by the Lagrangian
\begin{equation}
{\cal L}_{\rm EQT}={\cal L}_{H}+{\cal L}_F,
\end{equation}
where ${\cal L}_H$ is the Lagrangian of the Higgs sector and describes the
interactions of the four scalar particles $H,w^+,w^-$, and $z$ among each
other, and ${\cal L}_F$ is the fermionic Lagrangian which describes the Yukawa
interactions between the four scalar particles and the fermions of the theory.

In our investigation of the high energy behaviour of longitudinally polarized
gauge boson and Higgs boson scattering, we neglect the fermionic contributions
and concentrate on the physics determined by ${\cal L}_H$:
\begin{equation}
{\cal L}_{H}=\frac{1}{2}(\partial_\mu\Phi)^\dagger(\partial^\mu\Phi)
-\frac{\lambda}{4}(\Phi^\dagger\Phi)^2
+\frac{\mu^2}{2}(\Phi^\dagger\Phi).
\label{LSSB}
\end{equation}
Here $\Phi$ is a complex doublet.  After writing $\Phi$ in terms of real scalar
fields and introducing a vacuum expectation value (vev) for one of the fields,
we obtain the SM interactions:
\begin{eqnarray}
{\cal L}_{int} &=& 
-\; \frac{\lambda}{4}\left({\bf w}^4 + 2{\bf w}^2H^2 + H^4\right)
\;-\; \lambda v\left({\bf w}^2H + H^3\right),
\label{physlagr}
\end{eqnarray}
where ${\bf w}=(w^+,w^-,z)$.  The quartic interactions by themselves satisfy a
SO(4) symmetry, whereas the whole interaction Lagrangian is only SO(3)
symmetric~\cite{lytel}. These symmetries will reappear when discussing the
scattering amplitudes.

Note that ${\cal L}_{int}$ does not provide for trilinear couplings of the
Goldstone bosons.  Correspondingly, the trilinear gauge couplings of the
Standard Model are pure gauge couplings and are not subject of our analysis.

\section{OMS and $\overline\protect{\rm MS}$ renormalization}

The Lagrangian ${\cal L}_H$ must be renormalized under the constraint that the
Goldstone bosons remain massless at all orders of perturbation theory, i.e.,
the Goldstone theorem~\cite{ssb} applies.  Equivalently, we require the Higgs
field to be expanded around the minimum of the potential, acquiring a vacuum
expectation value of $v=2^{-1/4}G_F^{-1/2}\approx 246$~GeV. We use dimensional
regularization, so that the requirement above leads to the relation~\cite{mah}
\begin{equation}
\lambda_0=\frac{M_0^2}{Z_wM_H^2}\lambda.
\label{renfix}
\end{equation}
Here $\lambda_0$ ($\lambda$) is the bare (renormalized) quartic coupling, $Z_w$
is the field renormalization constant of the charged Goldstone bosons, and
$M_0$ ($M_H$) is the bare (renormalized) Higgs mass.  Note that
Eq.~(\ref{renfix}) is {\it renormalization scheme independent}.  In the limit
of zero Yukawa couplings, $Z_{w}\,=\,Z_z$ due to the SO(3) symmetry of ${\cal 
L}_{H}$.

In the OMS scheme, the explicit two-loop expressions for $\lambda_0$, $Z_w$,
and $M_0$ are given in \cite{mah}, and the necessary expressions for the
self-energies and wave-function renormalizations have also been reported in
\cite{ghi}.  In this scheme, the mass $M_H$ is defined as the pole mass, i.e.,
the physical mass of the Higgs boson. Counterterms are defined such that the
value of $M_H$ remains unchanged when going to higher orders in perturbation
theory. In other words, the pole of the Higgs propagator is always at the
physical mass $M_H$.  Similarly, the OMS scheme fixes the vacuum expectation
value $v$ to have the same value at each order. Hence, the tree level relation
\begin{equation}
\lambda_{\rm OMS}=\frac{M_H^2}{2v^2}
\label{fixcoup}
\end{equation}
is unchanged by higher-order corrections.  Therefore, the OMS-value of the
Higgs coupling, $\lambda_{\rm OMS}$, is given by Eq.~(\ref{fixcoup}) to {\it 
all orders} in perturbation theory.

In the $\overline{\rm MS}$ scheme, we find the following results:
\begin{equation} 
Z_w = Z_H = 1 +
\frac{\lambda_{\overline{\rm MS}}^2\;\xi^{2\epsilon}}{(16\pi^2)^2}
\Biggl(-\frac{3}{\epsilon}\Biggr)
+ {\rm O}(\lambda_{\overline{\rm MS}}^3),
\label{zwmsbar}
\end{equation}
\begin{eqnarray}
\lambda_0 \!&\!=& \lambda_{\overline{\rm MS}}\;\biggl[
1+\frac{\lambda_{\overline{\rm MS}}\;\xi^{\epsilon}}{16\pi^2}
\Biggl(\frac{12}{\epsilon}\Biggr)
+\frac{\lambda_{\overline{\rm MS}}^2\;\xi^{2\epsilon}}
 {(16\pi^2)^2}\Biggl(\frac{144}{\epsilon^2}
-\frac{78}{\epsilon}\Biggr) 
+ {\rm O}(\lambda_{\overline{\rm MS}}^3)\biggr],
\label{lambda0msbar}
\end{eqnarray}
\begin{eqnarray}
M_0^2 \!&\!=& \overline{M}^2\;\biggl[
1+\frac{\lambda_{\overline{\rm MS}}\;\xi^{\epsilon}}{16\pi^2}
\Biggl(\frac{12}{\epsilon}\Biggr)
+\frac{\lambda_{\overline{\rm MS}}^2\;\xi^{2\epsilon}}
 {(16\pi^2)^2}\Biggl(\frac{144}{\epsilon^2}
-\frac{81}{\epsilon}\Biggr) 
+ {\rm O}(\lambda_{\overline{\rm MS}}^3)\biggr],
\label{massmsbar}
\end{eqnarray}
where $\epsilon=(4-D)/2$, $D$ is the dimensionality of space-time, and
$\xi=4\pi e^{-\gamma_E}$, with $\gamma_E=0.5772\ldots$ the Euler
constant. 
The $\overline{\rm MS}$ Higgs mass is denoted as $\overline{M}$.

Knowing the bare coupling $\lambda_0$ in terms of both the OMS coupling and the
${\overline{\rm MS}}$ coupling, we can calculate the relation between
${\lambda}_{\overline{\rm MS}}$ and ${\lambda}_{\rm OMS}$ order by order in
perturbation theory. Up to two loops we find:
\begin{eqnarray}
\label{coup}
{\lambda}_{\overline{\rm MS}}(\mu_0)
={\lambda}_{\rm OMS}\Biggl[1\, \Biggr.
&+&\, \left( 12 \ln({\mu_0^2}/{M_H^2}) + 25 - 3\pi\sqrt{3}\right)
\frac{\lambda_{\rm OMS}}{16\pi^2}\, \nonumber \\
&+&\, \left( 144\ln^2({\mu_0^2}/{M_H^2})
+ \left(444 - 72\pi\sqrt{3}\right)\ln({\mu_0^2}/{M_H^2})
+ 524\zeta(2)\right.\nonumber\\
&&\left. - 90\zeta(3) 
+ 216\sqrt{3}{\bf Cl} - 225\pi\sqrt{3}  + 48\pi{\bf Cl} + 162K_5 + 75
\right)\frac{\lambda_{\rm OMS}^2}{(16\pi^2)^2} \nonumber\\
&+&\Biggl.\, {\rm O}\left({\lambda}_{\rm OMS}^3\right)
\Biggr],\\
\approx{\lambda}_{\rm OMS}\Biggl[1\, \Biggr.
&+&\, \left( 12 \ln({\mu_0^2}/{M_H^2}) + 8.676\right)
\frac{\lambda_{\rm OMS}}{16\pi^2}\, \nonumber \\
&+&\, \left( 144\ln^2({\mu_0^2}/{M_H^2})
+ 52.219\ln({\mu_0^2}/{M_H^2})
+ 286.836\right)\frac{\lambda_{\rm OMS}^2}{(16\pi^2)^2} \nonumber\\
&+&\Biggl.\, {\rm O}\left({\lambda}_{\rm OMS}^3\right)
\Biggr],
\end{eqnarray}
where $M_H$ is the physical Higgs mass, and $\mu_0$ is the arbitrary mass scale
in dimensional regularization.  The constant $K_5=0.92363\ldots$ was evaluated
numerically in \cite{mah}.  The Riemann $\zeta$ function takes the values
$\zeta(2)=\pi^2/6$ and $\zeta(3)=1.20205\ldots$, ${\bf Cl}$ is the maximum of
Clausen's function, ${\bf Cl}\equiv{\rm Cl}({\pi\over3})=1.01494\ldots$.  The
two-loop constant $(286.836\ldots)$ contains contributions from the one-loop
${\rm O}(\epsilon)$ term of the bare coupling in the OMS scheme.  Note that the
coefficients of the $n$-loop logarithmic terms, $\ln^m(\mu_0^2/M_H^2),\;1\leq
m\leq n$, can also be determined using the coefficients of the $n$-loop beta
function in connection with the ($n$-1)-loop result.  To one loop, our
expression agrees with the result by Sirlin and Zucchini~\cite{sirzuc}. For a
further discussion we refer to \cite{uli}.

Because the OMS-coupling is entirely fixed by the choice of the physical Higgs
mass, $M_H$, the ${\overline{\rm MS}}$ coupling is now entirely fixed by the
choice of the scale $\mu_0$ and the value of $M_H$.  We make the natural choice
$\mu_0=M_H$ when using the $\overline{\rm MS}$ coupling. This choice also
guarantees that all higher-order logarithmic terms do not contribute to the
correction.


\section{SCATTERING AMPLITUDES}

We are now able to carry out our analysis of $2\rightarrow 2$ scattering
processes involving longitudinally polarized gauge bosons and the Higgs boson
in the limit $\sqrt{s},M_H\gg M_W$. We neglect gauge and Yukawa couplings, and
use the EQT as explained above. First, we briefly review the exact EQT one-loop
results~\cite{daw2,joh}, and compare them with the
corresponding high-energy, $\sqrt{s}\gg M_H\gg M_W$, one-loop results obtained
by \cite{daw1,marc,djl}.  This establishes the range of validity of the
high-energy approximation. Next, we consider the limit of high-energy
scattering at two loops.  The latter calculation yields the information
necessary to carry out a RGE analysis up to next-to-next-to-leading logarithms,
NNLL.  We investigate both the OMS- and the ${\overline{\rm MS}}$-scheme.

\begin{figure}[tb]
\vspace*{30pt}
\centerline{
\epsfysize=3.0in \rotate[l]{\epsffile{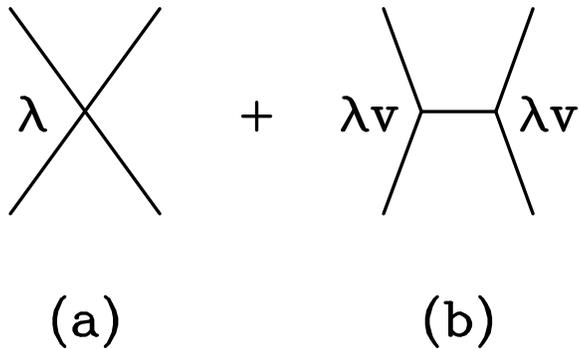}}
}
\vspace{0.15in}
\caption{The two topologies contributing to the $2\rightarrow2$ scattering
  processes at tree level.  For $\protect\sqrt{s}\protect\gtrsim 2M_H$ topology
  (a), the quartic coupling contribution to the amplitude, dominates relative
  to the trilinear coupling contributions of topology (b).  }
\label{feynman}
\end{figure}

The Feynman diagrams needed for the scattering processes are determined by the
interaction Lagrangian of Eq.~(\ref{physlagr}).  The relevant couplings are the
quartic coupling $\lambda$ and the trilinear coupling $\lambda v$, see
Fig.~\ref{feynman}.  We consider all possible two-body initial and final states
with total charge equal zero: $W_L^+W_L^-,\ Z_LZ_L,\ HH,\ Z_LH$; in the
notation of massless Goldstone bosons: $w^+ w^-,\ zz,\ HH,\ zH$.  Because of
the SO(3) symmetry of the Lagrangian ${\cal L}_H$, we can write the
unrenormalized transition amplitudes in terms of three different functions
$A',\,A'',\,A'''$, each depending on the Mandelstam variables $s,t,u$.  Taking
the two-body channels in the order $w^+ w^-,\ zz,\ HH,\ zH$, the $4\times4$
matrix of the unrenormalized transition amplitudes, \mbox{\boldmath${\cal F}$},
is given by~\cite{djl}
\begin{equation}
\mbox{\boldmath${\cal F}$} =
\left(
\begin{array}{cccc}
A'(s)+A'(t) & A'(s)                 & A''(s)                    & 0\\
A'(s)       & A'(s) + A'(t) + A'(u) & A''(s)                    & 0\\
A''(s)      & A''(s)                & A'''(s) + A'''(t)+A'''(u) & 0\\
0           & 0                     & 0                         & A''(t)
\end{array} \right),
\label{fmatrix}
\end{equation}
where we have indicated only the first variable in the functions $A$ since
these functions are unchanged by an interchange of the remaining two variables:
For example, $A'(t)\equiv A'(t,s,u)=A'(t,u,s)$.  The channels
$W_L^+W_L^-\rightarrow W_L^+W_L^-$ and $HZ_L\rightarrow HZ_L$ are the only
channels that do not display a $t\leftrightarrow u$ symmetry in their
amplitudes.

To obtain finite and physical $S$-matrix elements, we need to multiply the
unrenormalized amplitudes by the wavefunction renormalization constants of the
external fields, including finite parts such that the residue of the external
propagators is equal to unity.\footnote{In the OMS scheme this is done by
  definition. In the $\protect\overline{\rm MS}$ scheme, additional finite
  wavefunction renormalization constants need to be taken into
  account~\protect\cite{collinszhat}. } The physical transition amplitude is
then
\begin{equation}
\mbox{\boldmath${\cal M}$}={\bf Z}\mbox{\boldmath${\cal F}$}{\bf Z},
\label{feynmatrix}
\end{equation}
where $\bf Z$ is a diagonal matrix of renormalization constants,
\begin{equation}
{\bf Z}={\rm diag}(Z_w,Z_w,Z_H,Z_H^{1/2}Z_w^{1/2}).
\end{equation}
For a consistent calculation, the products in \mbox{\boldmath${\cal M}$} are to
be expanded to O($\lambda^3$), dropping higher orders.  

At high energies, $\sqrt{s}\gg M_H$, all internal particles of the scattering
diagrams can be taken massless. Hence, the Feynman diagrams only depend on
$\sqrt{s},\,\lambda,\,\lambda v$, and the scattering angle.  Because of the
even number of external particles, the trilinear coupling, $\lambda v$, must
always enter the $2\rightarrow 2$ Feynman diagrams in even powers. Furthermore,
$(\lambda v)^2 = M_H^2\lambda\,[1+{\rm O}(\lambda)] /2$ in any renormalization
scheme. Using dimensional arguments we conclude that Feynman scattering
diagrams involving trilinear couplings are suppressed by powers of $M_H^2/s$
relative to those which contain only quartic couplings.  To illustrate this
point, we look at the tree-level result for $W_L^+W_L^-\rightarrow Z_LZ_L$:
\begin{eqnarray}
A'(s,t,u) \;&=&\; -2\lambda \,-\, \frac{4\lambda^2v^2}{s-M_H^2} 
\,+\, {\rm O}(\lambda^2) \,+\, {\rm O}(\lambda^4v^4)\\
     &=&\; -2\lambda\left(1 + \frac{M_H^2}{s-M_H^2}\right) 
\,+\, {\rm O}(\lambda^2) \,+\, {\rm O}(\lambda^2 M_H^2)
\label{approx}\\
&\stackrel{s\gg M_H^2}{\rightarrow}&\; -2\lambda \,+\, {\rm O}(\lambda^2)
\,+\, {\rm O}\left(\frac{M_H^2}{s}\right).
\end{eqnarray}

Neglecting the trilinear couplings, the interaction Lagrangian,
Eq.~(\ref{physlagr}), becomes SO(4)-symmetric, and so do the high energy
Feynman amplitudes. That is,
\begin{equation}
\lim_{s\gg M_H^2}A'(s)\,=\,\lim_{s\gg M_H^2}A''(s)\,=\,\lim_{s\gg M_H^2}A'''(s)
\;\equiv\; A(s),
\end{equation}
thus simplifying the scattering amplitudes significantly.

The physical scattering amplitude \mbox{\boldmath${\cal M}$}, however, has no
SO(4) symmetry, since the renormalization constants $Z_w$ and $Z_H$ are defined
at the renormalization points $p^2=0$ and $p^2=M_H^2$, respectively. Therefore,
they contain contributions involving trilinear couplings, breaking the SO(4)
symmetry.  In addition, the Higgs mass has to be kept non-zero when calculating
self-energies and renormalization constants.  This introduces a logarithmic
dependence of the high-energy $S$-matrix elements on $M_H$, despite the fact
that the Higgs mass occuring inside Feynman diagrams is set to zero.

It is of interest to know for which energies $\sqrt{s}$ the high-energy
approximation can be used. Looking at Eq.~(\ref{approx}), we expect the
difference of the exact and high-energy results to be of the order
$M_H^2/(s-M_H^2)$. For example, the choice $\sqrt{s}\approx 3M_H$ is expected
to give an error of 10 to 15\% in magnitude. We examine whether this changes at
the one-loop level by comparing the exact one-loop EQT result with the
high-energy EQT result in the OMS scheme.

\subsection{OMS amplitudes}
The exact one-loop renormalized transition amplitudes
including both quartic and trilinear Higgs coupling contributions are taken
from Eqs.~(3.5)--(3.7) of \cite{joh} and were also calculated in \cite{daw2}.
The high energy result, $\sqrt{s}\gg 
M_H$, is obtained by dropping scattering diagrams involving the trilinear
coupling $\lambda v$ and setting the Higgs mass of internal Higgs propagators
equal to zero. The resulting high-energy amplitudes agree with the one-loop
high-energy results reported in \cite{daw1,marc,djl}.  We also present the
two-loop high-energy amplitudes which we calculate using the results
of~\cite{mah}.

Since the matrix elements depend on the scattering angle, we integrate out this
angular dependence and compare partial-wave projected $2\rightarrow2$
amplitudes for angular momentum $j$. They are defined by\cite{djl}
\begin{equation}
\label{pwpamp}
        {\bf a}_j^{if}(s)=\frac{N_iN_f}{32\pi} 
\left( \frac{4|\vec{p_i}||\vec{p_f}|}{s} 
        \right)^{1/2}\int_{-1}^{1}d(\cos\theta)\, \mbox{\boldmath${\cal
        M}$}^{if}(s,\cos\theta) 
        P_j(\cos\theta ).
\end{equation}
The momentum-dependent prefactor approaches unity for $\sqrt{s}\gg M_H$.  The
factors $N_i,N_f$ incorporate the symmetry factors which must be inserted for
each pair of identical particles in the initial and final state,
$N_i,N_f=1/\sqrt{2}$ for $zz,HH$, and $N_i,N_f=1$ for $w^+w^-,zH$.

To discuss the validity of the high-energy approximation, we explicitly state
the result for the channel $W_LW_L\rightarrow Z_LZ_L$.  The analytical result
of this specific channel is given in Appendix~\ref{angular}.  Numerical
evaluation yields the following OMS high-energy amplitude up to two loops:
\begin{eqnarray}
W_L^+W_L^-\rightarrow Z_LZ_L: & & {\rm OMS\;\; scheme}
\phantom{\Bigl( \Bigr)}\nonumber\\
{\bf a}_{j=0}(s)&=&\frac{N_iN_f}{32\pi} 
\left( \frac{4|\vec{p_i}||\vec{p_f}|}{s} 
        \right)^{1/2}
\left(-4\,{ \lambda_{\rm OMS}} \right)
\Bigglb[ \;\;1
\Biggrb.\nonumber\\
& &+\;  \left[ \! \,12\,
{\rm ln} \left( \! \,{\displaystyle \frac {{s}}{{ M_H}^{2}}}\,
 \!  \right)  - 21.32 - 25.13\,i\, 
\!  \right] \,\frac{ \lambda_{\rm OMS} }{16\pi^2}\nonumber\\
& &+\;  \left[ {\vrule height0.93em width0em depth0.93em} \right. \! \! 
  144\,{\rm ln}^2 \left( \! \,{\displaystyle \frac {{s}}{{
 M_H}^{2}}}\, \!  \right)  - (655.8 + 603.2\,i)\,{\rm ln} \left( \! \,
{\displaystyle \frac {{s}}{{ M_H}^{2}}}\, \!  \right)  \nonumber\\
& &\;\phantom{+[} \Bigglb. + 684.4 + 1097.0\, i\,
 \! \! \left. {\vrule height0.93em width0em depth0.93em} \right]
 \frac{\lambda_{\rm OMS}^2}{(16\pi^2)^2}\Biggrb].
\label{ampoms}
\end{eqnarray}
Notice that the tree-level high-energy amplitude is independent of $s$ since
the kinematical prefactor is evaluated to be one. The tree-level result is
therefore not an 
adequate description of the high-energy amplitude if using $\lambda_{\rm OMS}$.

\begin{figure}[tb]
\vspace*{30pt}
\centerline{
\epsfysize=3.0in \rotate[l]{\epsffile{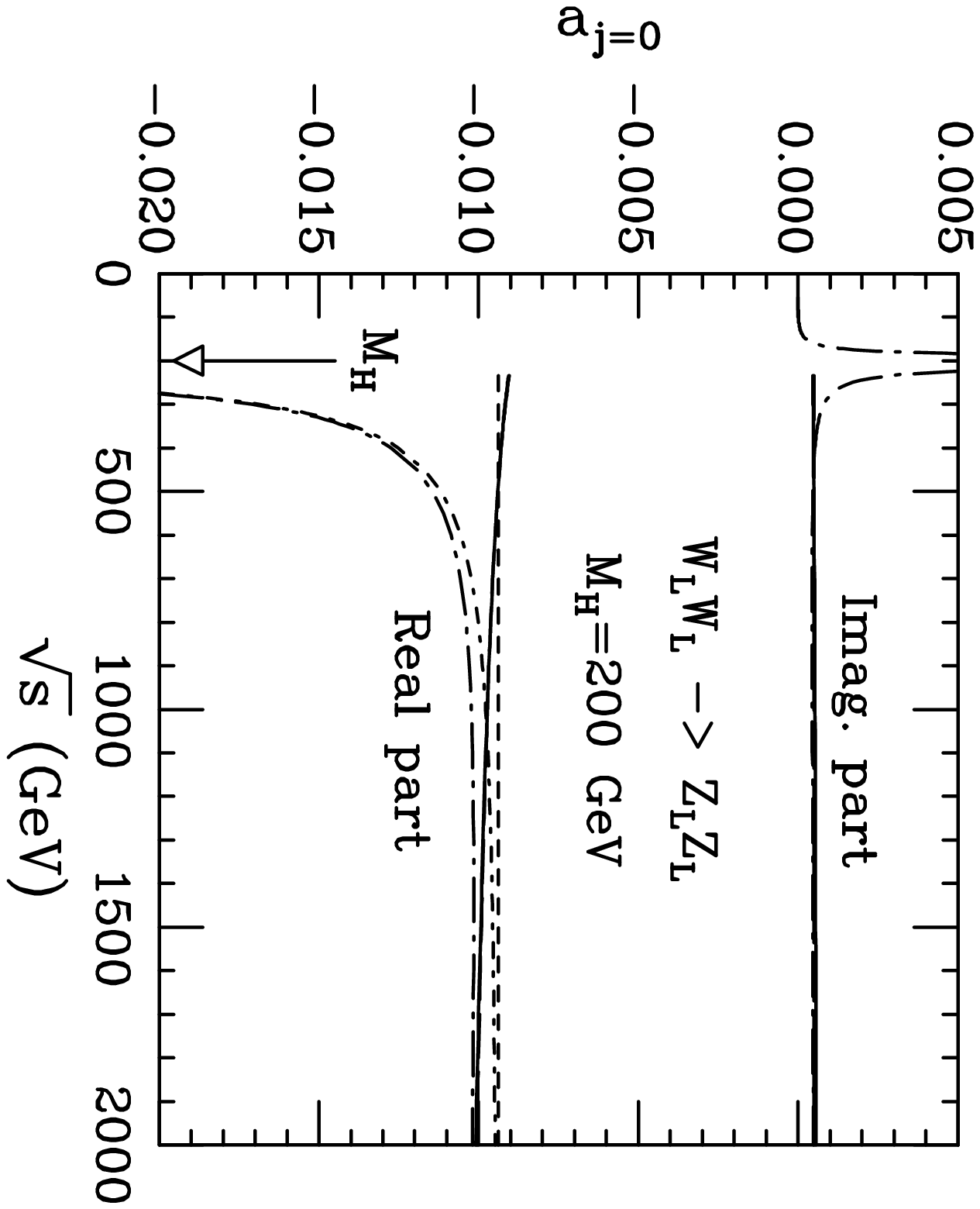}}
\epsfysize=2.86in \rotate[l]{\epsffile{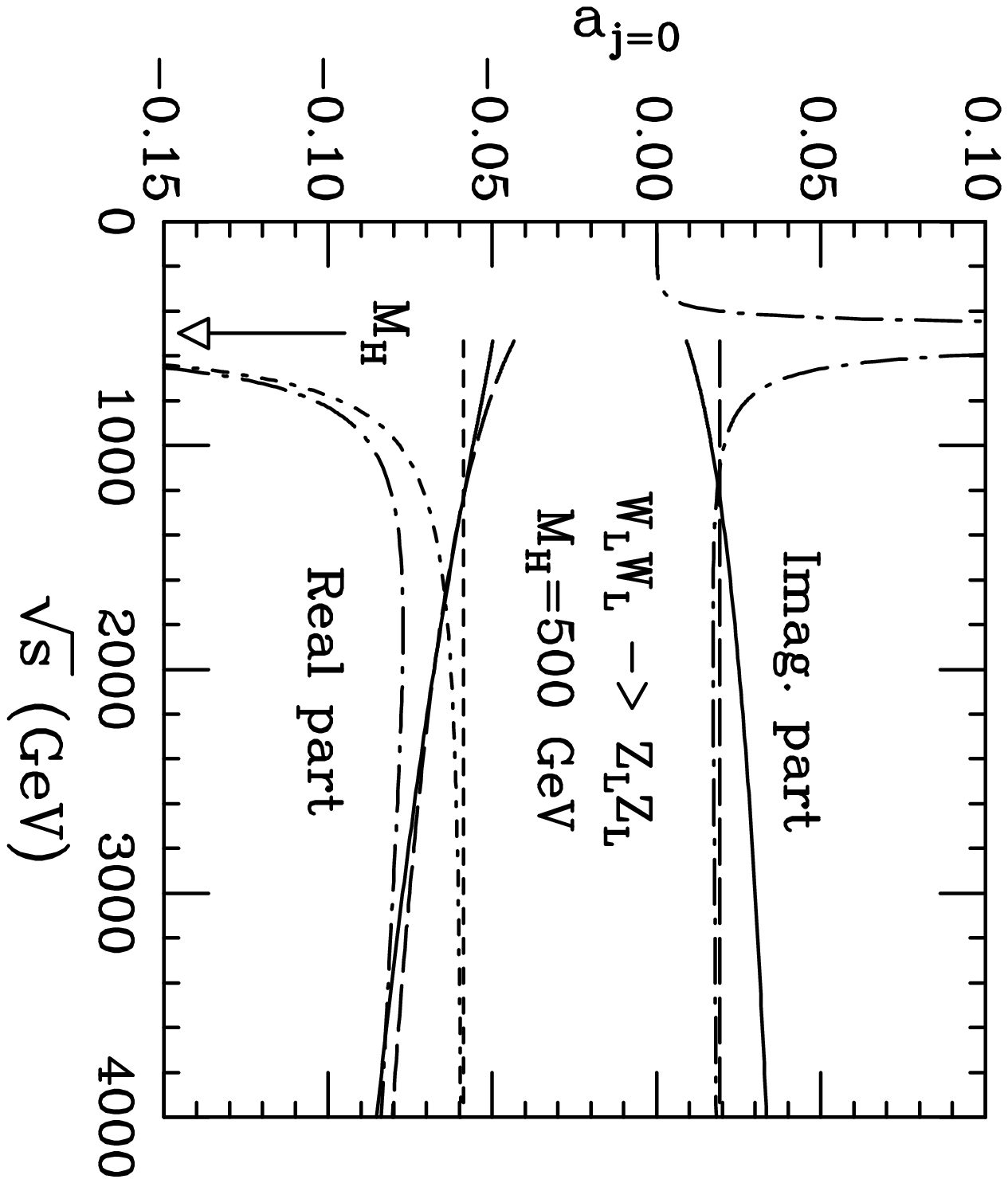}}
}
\vspace{0.15in}
\caption{The $j=0$ partial-wave projected amplitude for
  $W_LW_L\rightarrow Z_LZ_L$ in the limit of $M_H\gg M_W$. The Higgs mass is
  chosen as $M_H=200$ GeV (left plot) and $M_H=500$ GeV (right plot). The
  amplitudes are calculated in exact EQT approximation (dot-dashed curves,
  results taken from \protect\cite{daw2,joh}) and in high energy EQT
  approximation (dashed curves, \protect\cite{daw1,marc,djl}). Both tree-level
  (short dashes) and one-loop (long dashes) results are compared, and we find
  the high-energy approximation to differ less than 20\% for
  $\protect\sqrt{s}\protect\gtrsim 3M_H$.  We also show the two-loop high
  energy result (solid line).  Note the different scales used in the two plots.
  }
\label{comparison}
\end{figure}

In Fig.~\ref{comparison} we show the OMS $j=0$ partial-wave projected
amplitudes\footnote{The $2\rightarrow2$ processes considered here are
  predominantly $s$-wave processes, so that only $j=0$ is of interest.  For
  $W_L^+W_L^-\rightarrow Z_LZ_L$, the $j=1$ amplitude is zero, and the $j=2$
  amplitude is suppressed roughly by a factor of 100.} of the channel
$W_L^+W_L^-\rightarrow Z_LZ_L$, giving the high-energy EQT result at tree and
one-loop level, and comparing them with the corresponding tree and one-loop
results of the exact EQT calculation.  In addition, we also show the
high-energy two-loop result. These results are compared for both $M_H=200$
(left plot) and $500$ GeV (right plot). Note that the amplitude scale is a
factor 10 different in the two plots. This is necessary since the Higgs
coupling $\lambda_{\rm OMS}$ increases a {\it factor} of 6.25 when going from
$M_H=200$ to $M_H=500$ GeV.

We show ${\rm Re}\,a_{j=0}$ only for values $\sqrt{s}>M_H$. (For values of
$\sqrt{s}<M_H$, the real part becomes positive and is not shown.) The exact EQT
result features the typical pole~\cite{sey} at $\sqrt{s}=M_H$, whereas the
high-energy amplitude remains finite at the pole location. Increasing
$\sqrt{s}$, the exact result approaches the high-energy result rather quickly,
confirming the suppression of the trilinear coupling contributions at higher
energies. We find that for values $\sqrt{s}>3M_H$, the relative difference
\begin{equation}
\Delta(a_{j=0})=\frac{a_{j=0}^{\rm exact}-a_{j=0}^{\rm high}}
{a_{j=0}^{\rm exact}}
\end{equation}
between the exact and high-energy EQT result is less than 12\% at tree level.
However, taking higher order corrections into account, the relative difference
increases. The heavy-Higgs case shows a larger difference than the light-Higgs
case when keeping $\sqrt{s}/M_H$ fixed. This is due to the fact that a larger
Higgs mass also causes a larger Higgs coupling, increasing the importance of
higher-order corrections.  In Table~\ref{table1} we show the relative
differences for three different values of $M_H$ and a range of $\sqrt{s}$.
Choosing $M_H<500$ GeV the error induced by using the high-energy approximation
is less than approximately 20\% if $\sqrt{s}\gtrsim3M_H$. For such
cms-energies, the numerical difference between different orders in perturbation
theory (e.g.\ tree-level vs. one-loop) is much more significant than the
difference between exact EQT and high-energy EQT result.

It is also interesting to note that for $\sqrt{s}\approx1.5M_H$ the high-energy
result contributes about 50\% to the real part of the $j=0$ partial-wave
projected amplitude, i.e., the quartic coupling becomes dominant. Since it is
desirable to measure the quartic coupling $\lambda$ and the trilinear coupling
$\lambda v$ separately, the two different contributions to the cross section
need to be separated. This only seems feasible by going to the high-energy
region in which the trilinear coupling is completely suppressed.  Since the
cross sections decrease quickly for $\sqrt{s}> M_H$ this is a difficult task.
If the quartic coupling can be measured at high energies ($\sqrt{s}\approx 3-5
M_H$), one can extrapolate the high-energy cross section back to
$\sqrt{s}\approx M_H$. Subtracting the quartic coupling contribution from the
cross section at the resonance would yield the pure trilinear contribution at
the resonance.  To do the extrapolation from high energies to resonance
energies, the behaviour of the high-energy amplitude as a function of
$\sqrt{s}$ has to be well understood. This will be the subject of
Section~\ref{sect:rge}, where we introduce RGE methods.  Before doing so, we
briefly state the results of the high-energy amplitudes when using the
$\overline{\rm MS}$ scheme.

\subsection{$\protect\overline{\rm MS}$ AMPLITUDES}

The high-energy $\overline{\rm MS}$ amplitudes are calculated in two different
ways. One way to obtain the amplitudes is to calculate the high-energy Feynman
diagrams, taking all particles to be massless. The result is renormalized using
the  $\overline{\rm MS}$ definitions for the bare coupling and the wavefunction
renormalization constant, Eqs.~(\ref{zwmsbar}) and~(\ref{lambda0msbar}).
Finally, the physical transition amplitude is obtained by multiplying the
renormalized four-point functions with the finite renormalization constants of
the external fields~\cite{collinszhat}.

Alternatively, the $\overline{\rm MS}$ transition amplitudes can also be
calculated by taking the result of the OMS amplitudes and expressing the OMS
coupling in terms of the $\overline{\rm MS}$ coupling by means of
Eq.~(\ref{coup}).  The quantity $M_H$ appearing in the final amplitudes refers
to the pole mass of the Higgs propagator, {\it not} the $\overline{\rm MS}$
mass.

Again choosing the specific channel $W_LW_L\rightarrow Z_LZ_L$, the
$\overline{{\rm MS}}$ partial-wave projected amplitude in the high-energy limit
is:
\begin{eqnarray}
W_L^+W_L^-\rightarrow Z_LZ_L: & & \overline{\rm MS} \;\;{\rm
scheme}
\phantom{\Bigl( \Bigr)}\nonumber\\ 
{\bf a}_{j=0}(s)&=&\frac{N_iN_f}{32\pi} 
\left( \frac{4|\vec{p_i}||\vec{p_f}|}{s} 
        \right)^{1/2}
\Bigl( -4\,{\lambda}_{\overline{\rm MS}}(\mu_0) \Bigr)
\Bigglb[ \;\;1
\Biggrb.\nonumber\\
& &+\;  \left[ \! \,12\,
{\rm ln} \left( \! \,{\displaystyle \frac {{s}}{{ \mu_0}^{2}}}\,
 \!  \right)  - 30.000  - 25.13\,i\, 
\!  \right] \,\frac{ {\lambda}_{\overline{\rm MS}}(\mu_0)
}{16\pi^2}\nonumber\\ 
& &+\;  \left[ {\vrule height0.93em width0em depth0.93em} \right. \! \! 
  144\,{\rm ln}^2 \left( \! \,{\displaystyle \frac {{s}}{{
 \mu_0}^{2}}}\, \!  \right)  - (876.0 +603.2\,i)\,{\rm ln} \left( \! \,
{\displaystyle \frac {{s}}{{ \mu_0}^{2}}}\, \!  \right)  
+ 12\,{\rm ln} \left( \! \,
{\displaystyle \frac {s}{M_H^2}\, }\!  \right) 
\nonumber\\
& &\phantom{+[} \Bigglb. + 918.1 + 1533.1\, i
 \! \! \left. {\vrule height0.93em width0em depth0.93em} \right]
 \frac{[{\lambda}_{\overline{\rm MS}}(\mu_0)]^2}{(16\pi^2)^2}\;\:\Biggrb]\, .
\label{ampmsbar}
\end{eqnarray}
At each order, the leading terms in $\ln(s)$ have the same coefficients as in
the OMS scheme, see Eq.~(\ref{ampoms}). The difference is in the constant
terms which also lead to different coefficients for the subleading logarithms.
The scale $\mu_0$ is the scale at which the Higgs 
$\overline{\rm MS}$ coupling is defined, and its natural value is
of the order of the Higgs mass, $\mu_0\approx M_H$.  

The $\ln(s/M_H^2)$ term of the $\overline{\rm MS}$ result is due to the finite
wavefunction renormalization of the external fields which enter the physical
transition ampltitudes. The wavefunction renormalizations are low-energy
quanties.  At one-loop, $Z_w=Z_z$ is finite, and no $\ln(M_H)$ terms occur. At
two loops, $Z_w$ is divergent, and the finite pieces of the wavefunction
renormalization constants provide for a $\ln(M_H)$ dependence.

The result of the previous equation can be compared with the OMS amplitudes of
Eq.~(\ref{ampoms}).  We find that the $\overline{\rm MS}$ constants are larger
than the corresponding OMS quantities.  Evaluating the $\overline{\rm MS}$
coupling for $\mu_0=M_H$, we find that $\lambda_{\overline{\rm MS}}(M_H)$ is
larger than $\lambda_{\rm OMS}$. Hence {\it both} the coefficients {\it and}
the coupling are larger than the corresponding quantities of the OMS scheme,
resulting in larger radiative corrections.  A similar effect is observed in the
${\rm O}(\lambda^2)$ corrections in the Higgs decays $H\rightarrow f\bar
f$~\cite{willey,hffproc,uli} and $H\rightarrow W^+W^-$~\cite{uli}. A more
detailed discussion of the $\overline{\rm MS}$ 
amplitudes is provided at the end of the following section.

\section{RENORMALIZATION GROUP METHODS}
\label{sect:rge}

The previous section provided the amplitudes as obtained from calculating
Feynman diagrams to a certain order in perturbation theory.  Using
renormalization group techniques, we are also able to resum the energy
dependence of the amplitudes at higher orders. In the context of weak gauge
boson scattering this was originally introduced by \cite{daw1,marc,djl} at the
one-loop level.

The amplitudes are subject to renormalization group equations.  Using the OMS
scheme, the high-energy transition amplitude must satisfy the homogenous
Callan-Symanzik-equation~\cite{callan}:
\begin{equation}
\Biggl[ 
\;M_H\frac{\partial}{\partial M_H}
\;+\; \beta(\lambda_{\rm OMS})\frac{\partial}{\partial\lambda_{\rm OMS}}
\;-\; \sum_{i=1}^{4}\gamma_i
\;\Biggr]\; {\bf a}_{j} = 0.
\label{calsym}
\end{equation}
This equation is only valid in the high-energy region.  At low
energies, the right-hand-side of the equation is replaced by an inhomogenous
term. 

Using the $\overline{\rm MS}$ scheme, the transition amplitude must satisfy the
't~Hooft-Weinberg-equation~\cite{thooft}:
\begin{equation}
\Biggl[ 
\;\mu\frac{\partial}{\partial \mu}
\;+\; \overline{\beta}(\lambda_{\overline{{\rm
MS}}})\frac{\partial}{\partial\lambda_{\overline{{\rm
MS}}}} 
\;-\; \overline{\gamma}_{M}(\lambda_{\overline{{\rm
MS}}})\overline{M}\frac{\partial}{\partial \overline{M}}
\;\Biggr]\; {\bf a}_{j} = 0,
\label{weinberg}
\end{equation}
where $\overline{M}$ is the scale-dependent $\overline{\rm MS}$ mass.  This
differential equation is exact at all energies.  We will now solve the RGE and
discuss the RG improved amplitudes and cross sections in both schemes.

\subsection{RGE in OMS scheme}
The physical meaning of the homogenous Callan-Symanzik-equation,
Eq.~(\ref{calsym}), can be stated as follows: If all momenta are scaled by a
factor $\sigma $ so that $s,t,u\rightarrow\sigma ^2s,\sigma ^2t,\sigma ^2u$,
the scaled and original $2\rightarrow 2$ scattering amplitudes are related
by~\cite{chengli}
\begin{equation}
\label{ii6}
        {\cal M}\left( \left\{ \sigma p_i \right\},\lambda ,M_H 
        \right)=\Gamma_{i_1}\Gamma_{i_2}{\cal M}\left( \left\{ p_i 
        \right\},\lambda _s(\sigma),M_H \right)\Gamma_{f_1}\Gamma_{f_2}.
\end{equation}
The $\Gamma_\phi$ are related to the field anomalous dimensions of the Higgs
sector: 
\begin{equation}
\label{ii8}
      \Gamma _\phi={\rm exp}\left( -\int_{\lambda _s(1)}^{\lambda _s(\sigma 
      )}\frac{\gamma_\phi(\lambda )}{\beta (\lambda )}d\lambda  \right),\qquad 
        \phi=w^\pm,z,H,
\end{equation}
and $\lambda _s(\sigma )$ is the
OMS running coupling. The functions $\beta$ and  $\gamma_\phi$ are the
usual Callan-Symanzik renormalization group functions, in particular:
\begin{eqnarray}
\label{beta}
\beta(\lambda)\; &\equiv&\;\mu\frac{d\lambda}{d\mu}
\;\equiv\;\frac{\beta_0}{16\pi^2}\lambda^2
+\frac{\beta_1}{(16\pi^2)^2}\lambda^3
+\frac{\beta_2}{(16\pi^2)^2}\lambda^4
+{\rm O}\Bigl(\lambda^5\Bigr)\,.
\end{eqnarray}

The OMS coefficients of the $\beta$ function have been calculated to three
loops~\cite{beta,uli}: 
\begin{equation}
\label{betacoeff}
\beta_0=24\,,\qquad\beta_1\,=\,-\,312\,,
\qquad\beta_2^{\rm OMS}=4238.23\ldots\,.
\end{equation}
The values of $\beta_0$ and $\beta_1$ are scheme independent.  The three-loop
coefficient $\beta_2$ is scheme dependent, and its value given above refers to
the OMS scheme.  The field anomalous dimensions of the three Goldstone bosons
are identical and will be denoted as $\gamma_w$. At one loop, $\gamma_w$ and
$\gamma_H$ are zero.  At two loops, their value is scheme dependent.  In the
OMS scheme they are \cite{unit}
\begin{eqnarray}
\label{ii12}
        \gamma _w(\lambda)  
& = &  \gamma_{w,0} \frac{\lambda^2}{(16\pi^2)^2} + {\rm O}(\lambda^3),
       \quad \gamma_{w,0}=-6,\\
        \gamma _H(\lambda)  
& = & \gamma_{H,0} \frac{\lambda^2}{(16\pi^2)^2} 
        + {\rm O}(\lambda^3), \quad \gamma_{H,0}=(150-24\pi \sqrt{3})\,\approx
        19.41\ldots\, . 
\label{ii13}
\end{eqnarray}

The solution of the differential equation, Eq.~(\ref{beta}), defines the
running coupling $\lambda_s$.  Rewriting the parameter $\sigma$ as $\mu/\mu_0$
we have
\begin{equation}
  \ln(\mu/\mu_0) =\int_{\lambda _s(1)}^{\lambda _s(\mu/\mu_0 )}\frac{d\lambda 
        }{\beta (\lambda )}\, .
\end{equation}
This equation can be solved for $\lambda_s(\mu/\mu_0)$ iteratively, assuming
the expansion parameter $\lambda(\mu_0)$ to be small, i.e., perturbation theory
to be valid.  Then the differential equation of the running coupling can be
solved iteratively and the two-loop answer is
\begin{eqnarray}
{\lambda_s^{(2)}(\mu/\mu_0)}&=& 
\lambda_s(1)\, \left[
1 -\beta_0\hat\lambda_s(1)\ln(\frac{\mu}{\mu_0})
+ \frac{\beta_1}{\beta_0}\hat\lambda_s(1)
\ln\left( 1 -\beta_0\hat\lambda_s(1)\ln(\frac{\mu}{\mu_0})\right)
\right]^{-1}
\label{pert2lp}
\end{eqnarray}
where $\hat\lambda_s\equiv\lambda_s/(16\pi^2)$.
The superscript $(2)$ indicates that this expression is the two-loop running
coupling. The iterative solution for the three-loop running coupling
$\lambda_s^{(3)}$ is given in~\cite{uli}, together with a discussion of other
perturbative solutions for the running coupling.

\begin{figure}[tb]
\vspace*{30pt}
\centerline{
\epsfysize=3.0in \rotate[l]{\epsffile{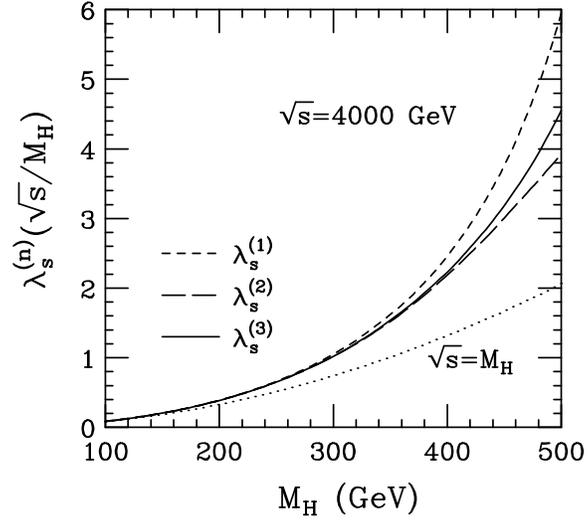}}
}
\vspace{0.15in}
\caption{
  The OMS running Higgs coupling $\lambda^{(n)}_s(\protect\sqrt{s}/M_H)$ in
  one-loop ($n$=1), two-loop ($n$=2), and three-loop ($n$=3) approximation. In
  all cases the running coupling is normalized at $\protect\sqrt{s}=M_H$ such
  that $\lambda^{(n)}_s(1)=M_H^2/(2v^2)$ (dotted curve).  Evolving the coupling
  to higher energies, the values of
  $\protect\lambda_s^{(1)},\lambda_s^{(2)},\lambda_s^{(3)}$ are shown for
  $\protect\sqrt{s}=4000$ GeV.  }
\label{elrun}
\end{figure}

The running coupling of a given renormalization scheme depends only on the
value of the Higgs quartic coupling at the scale $\mu=\mu_0$, i.e.
$\lambda_s(1)$. In the OMS scheme, we take $\mu_0=M_H$ and choose
$\lambda_s(1)$ to be equal to the non-running coupling $\lambda_{\rm OMS}$:
\begin{equation}
\label{ii11}
\lambda_s(1) =\lambda_{\rm OMS} = \frac{M_H^2}{2v^2}=\frac{G_\mu M_H^2}
{\sqrt{2}}\, .
\end{equation}
The evolution of the transition amplitude when going from the scale
$\mu=\mu_0\equiv M_H$ to the scale $\mu=sqrt{s}$ is now determined by the
evolution of the running coupling.  In Fig.~\ref{elrun} we show the running
coupling $\lambda_s(\sqrt{s}/M_H)$ at one, two, and three loops.  For each
value of $M_H$ we show the value of $\lambda_s(1)$, i.e, $\sqrt{s}=M_H$ (dotted
curve), and the value of the running coupling at $\sqrt{s}=4000$ GeV.
Requiring that the two-loop and one-loop coupling differ by less than 50\% for
$M_H\leq\sqrt{s}\leq 4000$ GeV, the Higgs mass should be less than 500
GeV.\footnote{For $M_H=500$ GeV the one-loop Landau pole of the one-loop
  coupling is located at $\sqrt{s}\approx 12$TeV.} As we will see below, this
value of $M_H$ is still too large for a perturbative RG treatment of
high-energy amplitudes and cross sections.

We are now able to estimate the importance the logarithms.  If $\sqrt{s}\gg
M_H$ they are expected to be very important. The question is: When is
$\sqrt{s}$ large?  To get a first estimate we look at the denominator of the
running coupling, Eq.~(\ref{pert2lp}), at one loop.  
With the above choices of $\mu$ and $\mu_0$, it is given by $1 -
24\ln(\sqrt{s}/M_H)\,M_H^2/(32\pi^2v^2).$ We expect to approach the
nonperturbative region if this quantity is about $0.5\,$; in other words, if
the one-loop running coupling $\lambda_s(\sqrt{s}/M_H)$ is twice the expansion
parameter $\lambda_s(1)$. For $M_H=200$ GeV we reach this limit if
$\sqrt{s}\approx 4\times 10^6$ GeV.  However, choosing $M_H=500$ GeV we enter
the nonperturbative region for $\sqrt{s}\approx 2500$ GeV.  This is already
within the $\sqrt{s}$ region considered previously, and the next-to-leading
logarithms will be important --- if perturbation theory doesn't fail at all.

{\bf OMS amplitudes:} Introduction of the running coupling
$\lambda_s(\sqrt{s}/M_H)$ into the transition amplitude \mbox{\boldmath${\cal
    M}$} results in a resummation of $\ln(s/M_H^2)$ terms.  The one-loop
running coupling, $\lambda_s^{(1)}$, resums terms of ${\rm
  O}\left(\lambda^{n+1}\ln^{n}(s/M_H^2)\right)$, the leading logarithms (LL).
The LL amplitude is hence obtained by the tree-level result of the amplitude in
connection with the one-loop running coupling.  
The LL amplitude therefore depends on $s$ --- in contrast to the naive
tree-level amplitude using $\lambda_{\rm OMS}$.
The
next-to-leading-log (NLL) amplitude corresponds to the one-loop amplitude using
a two-loop 
running coupling. This resums contributions ${\rm
  O}(\lambda^{n+2}\ln^{n}(s/M_H^2))$ to the amplitude to all orders.  Finally,
  the 
next-to-next-to-leading-log amplitude (NNLL) is obtained by including
one more loop in both the amplitude and the running coupling.  To clearify
this, we give the RG relations between coefficients occuring in the amplitudes
and the coefficients of the beta function in Appendix A, using as an example
the process $W_LW_L\rightarrow Z_LZ_L$.

Using the running coupling, there is always the question how the scale $\mu$ is
to be chosen. To resum the complete logarithmic dependence of the amplitude (or
cross section), one chooses $\mu=\sqrt{s}$ (and $\mu_0=M_H$ as stated earlier).
Of course, there is the possibility to choose $\mu$ different from
$\sqrt{s}$. Presently, there is no physical motivation which would suggest not
to resum the complete logarithmic dependence. Therefore we take
$\mu=\sqrt{s}$ throughout this paper.  For a discussion on the dependence of
the cross section on the choice of $\mu$ we refer to \cite{uli}.

Using the perturbative OMS amplitude of the previous section,
Eq.~(\ref{ampoms}), we obtain the NNLL amplitude of the channel
$W_L^+W_L^-\rightarrow Z_LZ_L$ in which the complete logarithmic dependence has
been resummed ($\mu=\sqrt{s}$):
\begin{eqnarray}
W_L^+W_L^-\rightarrow Z_LZ_L: & & \quad {\rm OMS\;\; with\;\; running\;\;
coupling}\nonumber\\ 
{\bf a}^{\rm NNLL}_{j=0}(s)\;\;=&&\frac{N_iN_f}{32\pi} 
\left( \frac{4|\vec{p_i}||\vec{p_f}|}{s} \right)^{1/2}
\Gamma_w^4\;
\Bigl({ -4}\,{ \lambda_s^{(3)}(\sqrt{s}/M_H)}\Bigr)
\Biggl[\;\; 1 \Biggr.\nonumber\\
& &-\;  \Bigl(\,  21.32 + 25.13\,i\, 
  \Bigr) \,\frac{ \lambda_s^{(3)}(\sqrt{s}/M_H)
}{16\pi^2}
{\vrule height0.93em width0em depth0.93em} \nonumber\\ 
& &+\; \Biggl.  \Bigl( 684.4 + 1097.0\, i\,
 \!  \Bigr)
 \frac{\left[\lambda_s^{(3)}(\sqrt{s}/M_H)\right]^2}{(16\pi^2)^2}\;\Biggr].
\label{wwzzamp}
\end{eqnarray}
The results for the other relevant $2\rightarrow 2$ scattering amplitudes of
the Higgs sector are given in
Appendix~\ref{resj0}.  Their relative sizes are very similar to the
$W_L^+W_L^-\rightarrow Z_LZ_L$ channel discussed.

In Eq.~(\ref{wwzzamp}) the factor $\Gamma_w$ depends on the anomalous dimension
of the Goldstone 
fields and resums a small residual dependence on $\ln(s/M_H^2)$ at two loops.
Using Eq.~(\ref{ii8}), $\Gamma_w$ is given by~\cite{mah,chengli} 
\begin{eqnarray}
\Gamma_w = \exp\biggl(\frac{-\gamma_{w,0}}{\beta_0}\;
           \frac{\lambda_s(\sqrt{s}/M_H)-\lambda_s(1)}{16\pi^2}\biggr)
\approx\left(\frac{\sqrt{s}}{M_H}\right)^{\frac
           {-\gamma_{w,0}\lambda_s^2}{16\pi^2}}\, . 
\end{eqnarray}
For channels involving external Higgs fields, we also need
\begin{eqnarray}
\Gamma_H = \exp\biggl(\frac{-\gamma_{H,0}}{\beta_0}\;
           \frac{\lambda_s(\sqrt{s}/M_H)-\lambda_s(1)}{16\pi^2}\biggr)
\approx \left(\frac{\sqrt{s}}{M_H}\right)^{\frac
           {-\gamma_{H,0}\lambda_s^2}{16\pi^2}}\, . 
\end{eqnarray}
Because of the smallness of the anomalous dimensions, Eqs.~(\ref{ii12}),
(\ref{ii13}), these factors are very close to unity: If $M_H\leq 500$ GeV and
$M_H\leq\sqrt{s}\leq 4000$ GeV then $0.980<\Gamma_H<1$ and $1<\Gamma_w<1.006$,
where the largest difference from unity corresponds to the largest values of
$M_H$ and $\sqrt{s}$.

The NLL amplitude is obtained from Eq.~(\ref{wwzzamp}) by dropping the two-loop
correction terms and 
using $\lambda_s^{(2)}$. In addition, the factors $\Gamma_w$ and $\Gamma_H$ 
are unity since the anomalous dimensions vanish at one loop.

Looking at Eq.~(\ref{wwzzamp}), it is interesting to note that the magnitude of
the ratio of one-loop to tree-level coefficient is about
$20/(16\pi^2)\approx0.13$, and the two-loop to one-loop ratio is roughly
$40/(16\pi^2)\approx0.25$ in magnitude. Since the running coupling is larger
than one for a large range of Higgs masses (see Fig.~\ref{elrun}) the
perturbative character of the series is already doubtful.

\begin{figure}[tb]
\vspace*{30pt}
\centerline{
\epsfysize=3.0in \rotate[l]{\epsffile{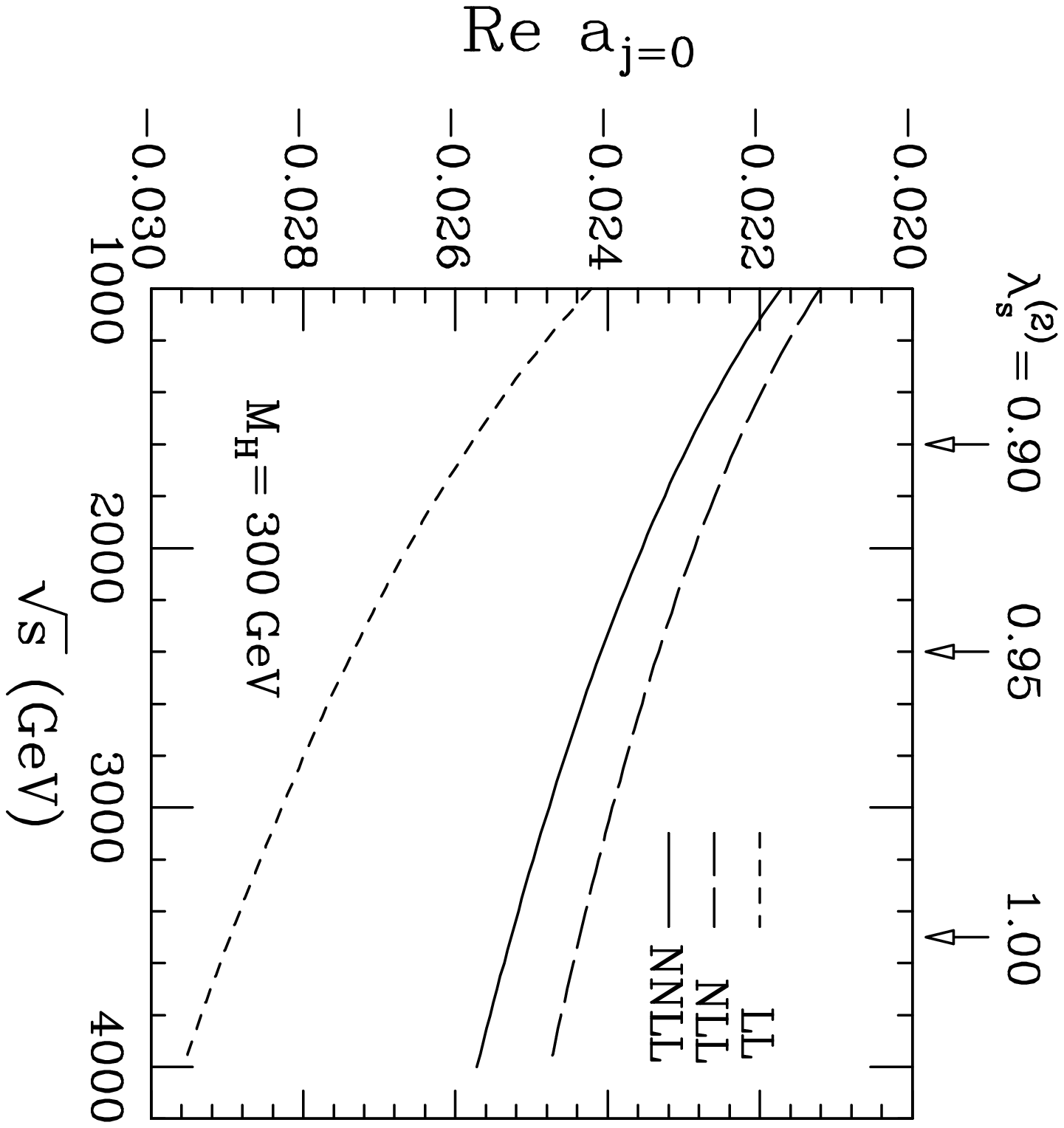}}
\hspace{.4cm}
\epsfysize=2.94in \rotate[l]{\epsffile{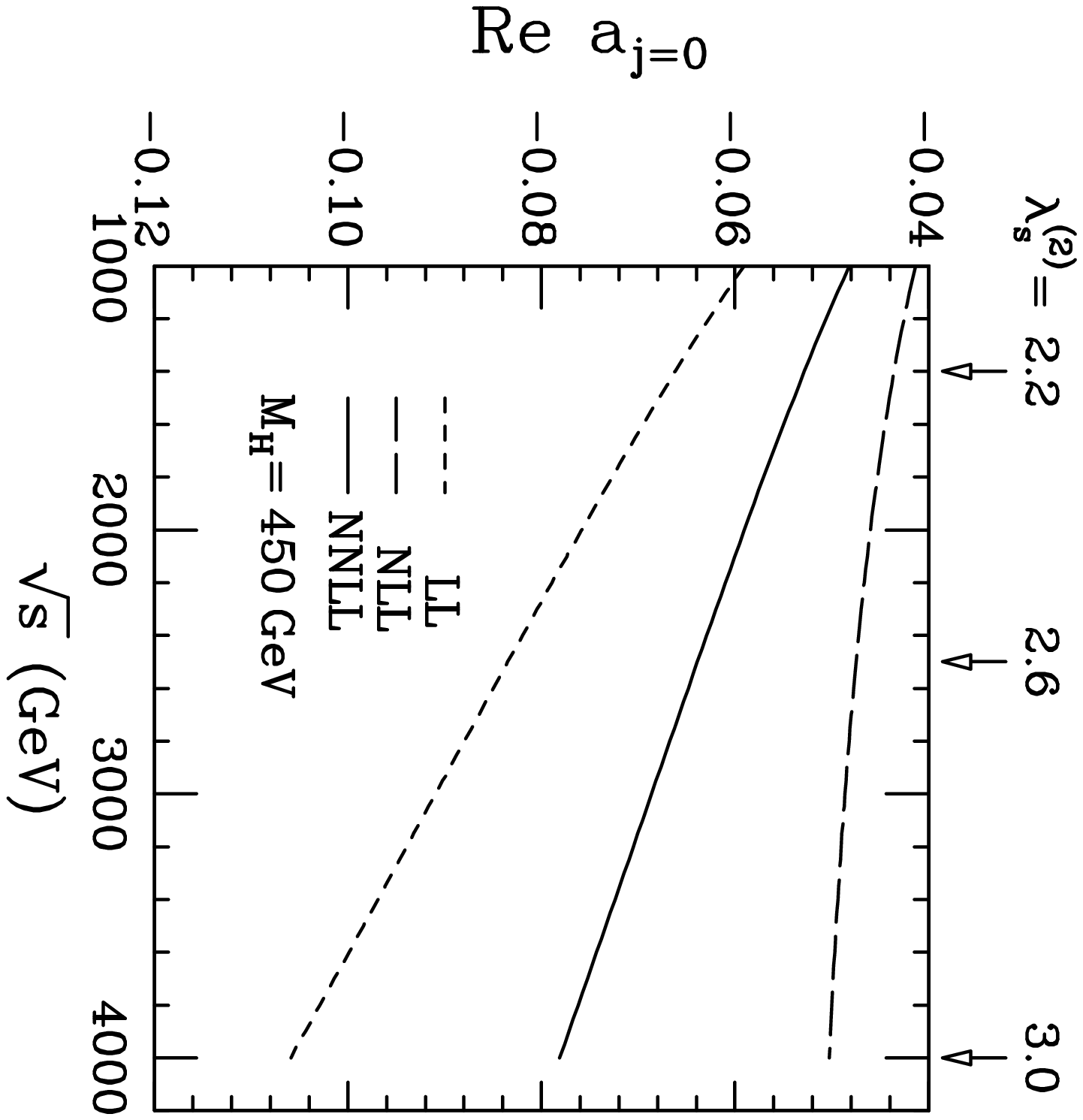}}
}
\vspace{0.15in}
\caption{The real part of the $j\!=\!0$ partial-wave projected amplitude for
  the process $W_L^+W_L^-\rightarrow Z_LZ_L$ for $M_H=300$ GeV (left plot) and
  $M_H=450$ GeV (right plot). The amplitudes are calculated using the
  high-energy approximation. At each order of the perturbative calculation, the
  complete $\ln(s/M_H^2)$ dependence of the amplitude is resummed using the
  $(n+1)\,$-loop running coupling $\lambda_s^{(n+1)}$ in connection with the
  $n$-loop perturbative amplitude.  This yields the LL, NLL, and NNLL results
  for $n=1,2,3$, respectively.  Note the very different scales used in the two
  plots.  }
\label{runa0}
\end{figure}

It is clear that the LL result can only provide for a first --- possibly
excellent --- estimate of the amplitude.  In Fig.~\ref{runa0} we discuss the
importance of the NLL and NNLL corrections for the high-energy amplitudes.  We
show the real part of the $j=0$ partial-wave projected amplitude of the channel
$W_L^+W_L^-\rightarrow Z_LZ_L$ in various approximations: LL, NLL, and NNLL.
The left plot shows the result for $M_H=300$ GeV. At $\sqrt{s}=1000$ GeV, the
magnitude of the LL result is reduced by about 10\% when going from LL to NLL
approximation.  At $\sqrt{s}=4000$ GeV, the reduction amounts to 20\%. The
additional correction from the NNLL calculation is at most a few percent for
the whole range of $\sqrt{s}$ considered.  The perturbative amplitude shows a
nice convergence.  The right plot of Fig.~\ref{runa0} shows the same quantity
taking $M_H=450$ GeV. The larger Higgs mass causes a larger running coupling,
leading to a larger amplitude and larger corrections. At $\sqrt{s}=1000$ GeV,
the LL result is reduced by 30\% when including the NLL corrections, and at
$\sqrt{s}=4000$ GeV the NLL corrections are even 50\%. The NNLL are important
for the whole range of $\sqrt{s}$ shown and also need to be included when
discussing cross sections.

{\bf OMS cross sections:} We now investigate the impact of our previous
findings when calculating physical observables, i.e., cross
sections.\footnote{Re $a_0$ is not a measurable quantity. Only $|a_0|^2$ is
  measurable, and it is closely related to the cross section discussed here.}
  We write the 
transition amplitude $\mbox{\boldmath${\cal M}$}^{if}$ of the scattering
process as
\begin{equation}
\mbox{\boldmath${\cal M}$}^{if}={\cal M}^{(0)}\,\lambda_s +
{\cal M}^{(1)}\,\lambda_s^2 + {\cal M}^{(2)}\,\lambda_s^3 + \ldots\, ,
\end{equation}
where ${\cal M}^{(0)}$ is real.
The  two-loop perturbative cross section in the cm system is
\begin{eqnarray}
\sigma_{if} =
\frac{N_i^2\,N_f^2}{32\pi s}\frac{ |\vec{p_f}|}{|\vec{p_i}|}
\; \Gamma_i^4\Gamma_f^4\;&&
\int_{-1}^{1} d\cos\theta  
\left[\rule{0pt}{10pt}\right.  ({\cal M}^{(0)})^2\,\lambda_s^2 \;+\;
2{\cal M}^{(0)}{\rm Re}\,{\cal M}^{(1)}\,\lambda_s^3 \nonumber\\
&&\left.\;+\;\rule{0pt}{10pt}
\Bigl(2{\cal M}^{(0)}{\rm Re}\,{\cal M}^{(2)} +  ({\rm Re}\,{\cal M}^{(1)})^2
+  ({\rm Im}\,{\cal M}^{(1)})^2 \,\Bigr)\lambda_s^4 + {\rm O}(\lambda_s^5) 
 \right]\, .
\end{eqnarray}
The indices $i,f=$ label the four possible neutral two-body states for the
$2\rightarrow 2$ process $i\rightarrow f$. We use the resummed form of the
transition amplitude, i.e., the logarithms are almost completely absorbed in
the running coupling $\lambda_s$, and a small two-loop logarithmic dependence
is resummed using $\Gamma_i$ and $\Gamma_f$.  To have a perturbatively
consistent two-loop cross section, we drop terms of ${\rm
  O}\left([\lambda_s]^5\right)$ since they get additional contributions from
diagrams involving three loops.  The one-loop cross section is obtained by
dropping all ${\rm O}\left([\lambda_s]^4\right)$ terms and setting
$\Gamma_i=\Gamma_f=1$.

We now discuss the total cross section of the channel $W_L^+W_L^-\rightarrow
Z_LZ_L$. Using the analytical two-loop results of Appendix~\ref{angular}, the
perturbative result of the high-energy resummed cross section is
\begin{eqnarray}
\sigma_{\rm OMS}^{\rm NNLL}\, 
& = & \frac{N_i^2\,N_f^2}{4\pi s}\frac{
|\vec{p_f}|}{|\vec{p_i}|}\;\Gamma_w^8\; 
\,\left[\lambda_s^{(3)}\right]^2\,
\Biggl[\,1
-\, 42.65\,\frac{\lambda_s^{(3)}}{16\pi^2}\,
+\; 2\,457.9\,\frac{\left[\lambda_s^{(3)}\right]^2}{(16\pi^2)^2}\,
+\; {\rm O}\left([\lambda_s^{(3)}]^3\right)\,\Biggr].
\label{wwzzcross}
\end{eqnarray}
The magnitude of the ratio of one-loop to tree-level coefficient is about
$40/(16\pi^2)\,\approx\,0.25$, and the ratio of two- to one-loop coefficient is
about $60/(16\pi^2)\,\approx\,0.4\,$: The convergence of the cross section is
apparently worse than the convergence of the amplitudes, Eq.~(\ref{wwzzamp}).
The reason is the fact that the one-loop correction of the cross section is
enhanced by a factor  two relative to the one-loop correction of the real
part of the amplitude. Hence the NLL corrections should always be included in
the cross section. At two loops, the squares of the one-loop real and
imaginary part also contribute to the two-loop correction, adding in magnitude
to other two-loop contributions. Depending on the value of the running
coupling, the NNLL correction may be very large.

\begin{figure}[tb]
\vspace*{30pt}
\centerline{
\epsfysize=3.0in \rotate[l]{\epsffile{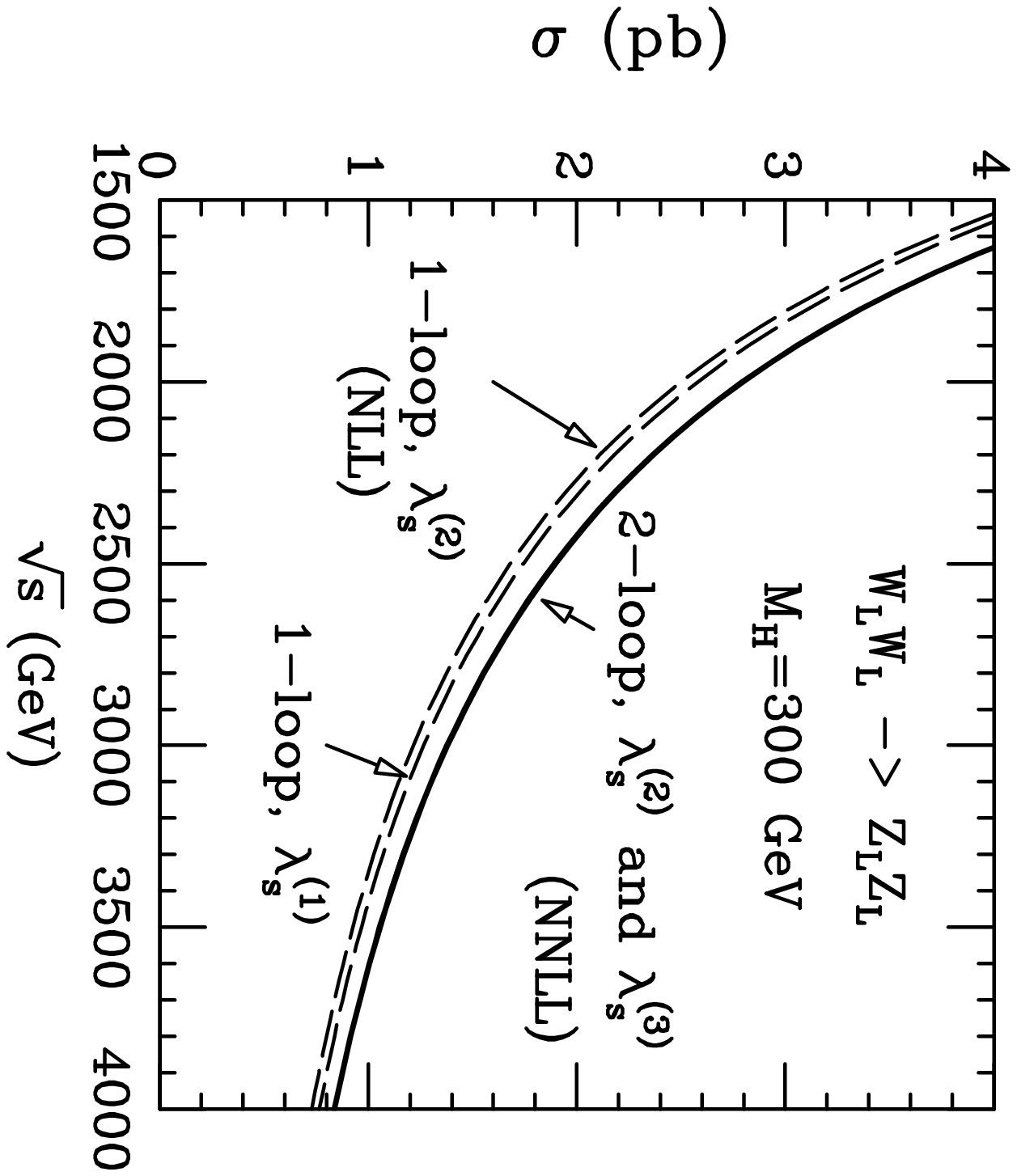}}
\hspace{.4cm}
\epsfysize=3.0in \rotate[l]{\epsffile{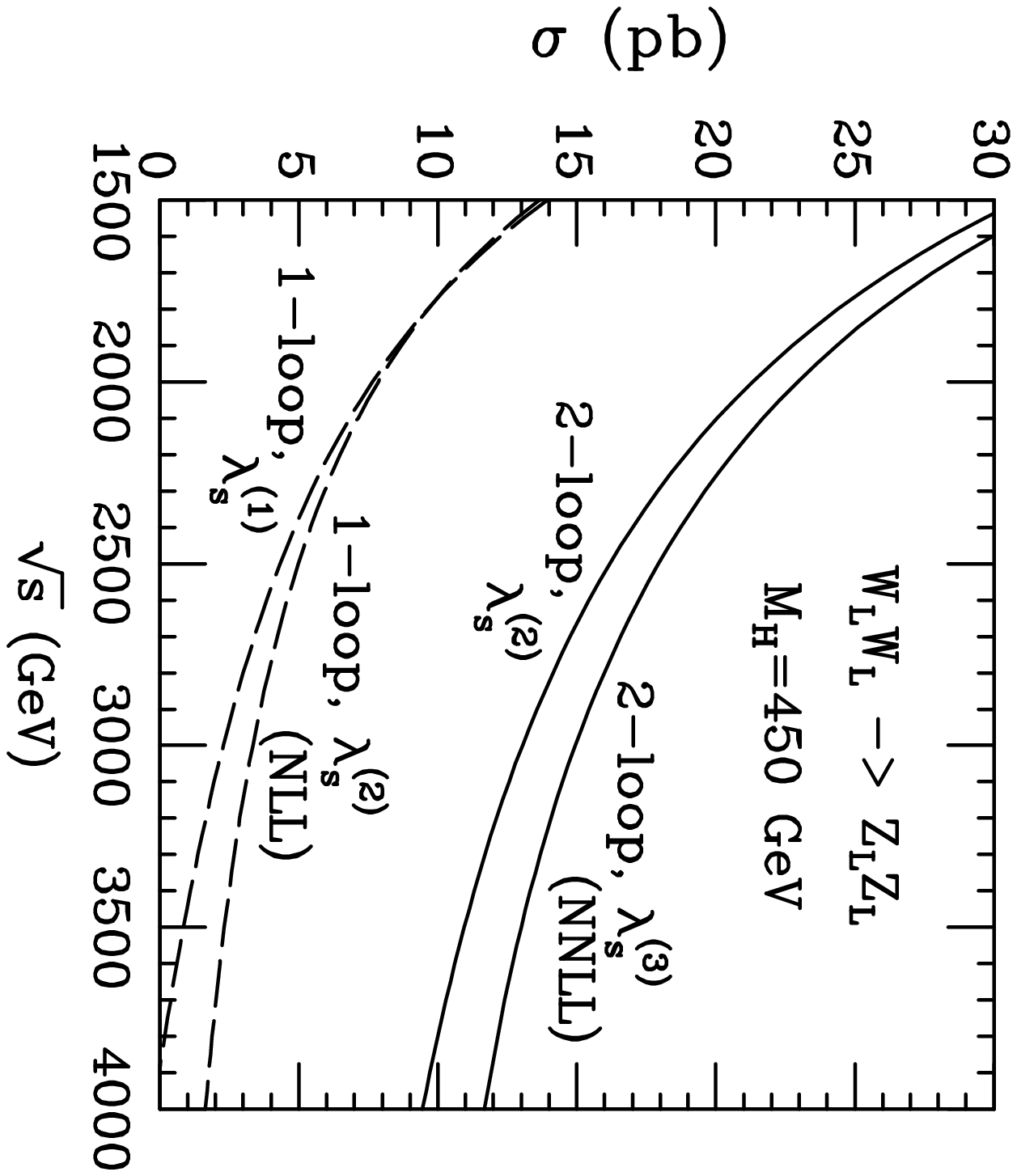}}
}
\vspace{0.15in}
\caption
{The cross section of the process $\protect W_L^+W_L^-\rightarrow Z_LZ_L$ as a
  function of the cms-scattering energy $\protect\sqrt{s}$ for $M_H=300$ GeV
  (left plot) and $450$ GeV (right plot).  The relative importance of the
  logarithmic and non-logarithmic contributions is examined at the one-loop and
  two-loop level.  Note that the one-loop perturbative cross section in
  connection with the one-loop running coupling is negative for $M_H=450$ GeV
  and $\protect\sqrt{s}\geq 3870$ GeV.  }
\label{sigmaoms}
\end{figure}

In Fig.~\ref{sigmaoms} we show the cross section for the process
$W_L^+W_L^-\rightarrow Z_LZ_L$ as a function of the cms-energy, choosing the
Higgs mass to be $M_H=300$ GeV (left plot) and $M_H=450$ GeV (right plot).  To
discuss the impact of resumming logarithmic terms at varies levels, we show the
one-loop cross section using both one-loop and two-loop running coupling, the
latter yielding the properly resummed NLL cross section. The first
combination does not give a full resummation of NLL terms, i.e. ${\rm
  O}\left(\lambda_s^{n+2}\ln^n(s/M_H^2)\right)$.  The two-loop cross section is
also given at two different levels of resummation: In connection with the
two-loop running 
coupling not all NNLL terms are resummed, whereas the use of the three-loop
running coupling gives the completely resummed NNLL cross section.

For $M_H=300$ GeV (left plot), the four different results are very similar,
with the one-loop results differing less than 20 percent from the two-loop
results for the whole range of energies considered.  More importantly, at a
given order it is not very important whether the full set of higher-order
logarithmic terms is resummed or not.  In particular, the two different
two-loop approximations yield almost identical results.

For $M_H=450$ GeV (right plot), the situation is very different.  The two-loop
results are much larger than the one-loop results.  At $\sqrt{s}=1500$ GeV, the
two-loop cross section is a about a factor two larger than the one-loop cross
section.  The difference between the two orders of calculation is even larger
for higher energies.  It is interesting that the level of resummation of
higher-order logarithmic terms is still much less important than the order of
the calculation. It is the non-logarithmic terms which are the cause of the
large corrections, and their importance increases for increasing Higgs
coupling, that is, increasing $\sqrt{s}$ and/or $M_H$.  For large values of
$\sqrt{s}$ and/or $M_H$ perturbation breaks down.  This is documented most
strikingly by the one-loop perturbative cross section which can become negative
since the one-loop correction to the perturbative cross section has the
opposite sign of the tree-level term; see Eq.~(\ref{wwzzcross}).\footnote{The
  perturbative cross section is not positiv definite since higher-order terms
  are dropped when squaring the amplitude.} For $M_H=450$ GeV and using the
one-loop running coupling, the one-loop cross section is negative for
$\sqrt{s}\gtrsim 3870$\ GeV, a clear signal for the breakdown of perturbation
theory.  Looking at Eq.~(\ref{wwzzcross}) we find that the one-loop
perturbative cross section becomes negative if $\lambda_s\geq 16\pi^2/42.65
\approx 3.7$, independent of a particular choice of $M_H$ and $\sqrt{s}$.  For
$M_H=300\;(450)$ GeV, the one-loop running coupling $\lambda_s^{(1)}$ is larger
than 3.7 if $\sqrt{s}\gtrsim 3.5\times 10^5 \;(3870)$ GeV.  Using the two-loop
running coupling $\lambda_s^{(2)}$, the one-loop NLL cross section can
also be negative.  Requiring $\lambda_s^{(2)}\gtrsim 3.7$ the one-loop NLL
cross section becomes negative for $\sqrt{s}\gtrsim 7800$\ GeV if $M_H=450$.

\begin{figure}[tb]
\vspace*{30pt}
\centerline{
\epsfysize=3.0in \rotate[l]{\epsffile{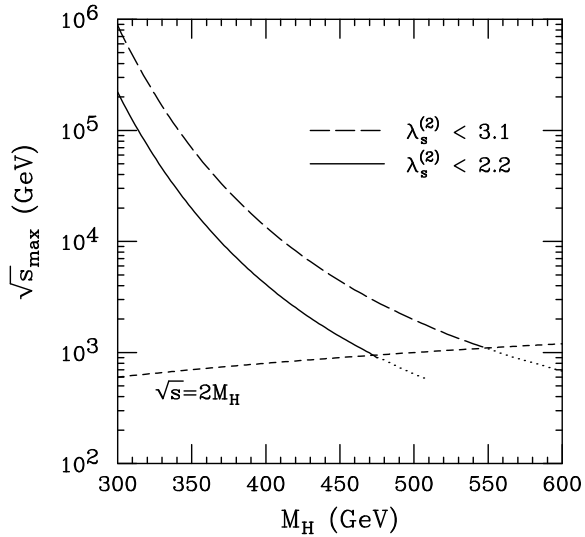}}
}
\vspace{0.15in}
\caption{
  The maximal cms energy $\protect\sqrt{s}$ up to which a perturbative
  calculation of $2\rightarrow 2$ scattering processes in the Higgs sector is
  possible. The upper bound on $\protect\sqrt{s}$ is derived from
  $\lambda_s(\protect\sqrt{s}/M_H)< 3.1$ which corresponds to requiring the
  one-loop RG improved $2\rightarrow 2$ cross sections to be positive. For
  larger values of $\protect\sqrt{s}$ (and hence $\lambda_s$) perturbation
  theory completely fails.  The stricter bound $\lambda_s<2.2$ is derived
  requiring the ratio $\sigma^{\rm NNLL}/\sigma^{\rm NLL}$ to be less than two
  in the OMS scheme, and it leads to a stricter upper bound on
  $\protect\sqrt{s}$. .  Values of $2.2\protect\lesssim\lambda_s<3.1$ lead to
  very unreliable perturbative cross sections.  The bounds on
  $\protect\sqrt{s}$ are calculated using the two-loop running coupling.  For
  the high-energy bounds to be applicable, we need $\protect\sqrt{s}>2M_H$. }
\label{limits}
\end{figure}

Our findings based on the above cross sections indicate that it is {\it
impossible} to make a satisfactory perturbative prediction for the high-energy
cross section for $W_L^+W_L^-\rightarrow Z_LZ_L$ 
from perturbation theory if $\lambda_s\approx 3.7$. Examining the cross
sections of the other $2\rightarrow 2$ channels of our scattering matrix, we
find critical values in the range of $\lambda_s=3.1$ to $4.1$, depending on the
size of the one-loop coefficient.  To obtain reliable perturbative cross
sections, the minimal requirement is 
\begin{equation}
\lambda_s < 3.1\,.
\label{limit1}
\end{equation}  
Keeping in mind that
the non-running coupling $\lambda_{\rm OMS}\approx 3$ for $M_H=600$ GeV and
that 
$\lambda_s>\lambda_{\rm OMS}$ for $\sqrt{s}>M_H$, the requirement $\lambda_s<
3.1$ is already a strong constraint on a perturbative Higgs mass if we want to
predict high-energy cross 
sections for $\sqrt{s}\gtrsim 2M_H$. 
To find out how much smaller than 3.1 the running coupling has to be, we
consider the case 
$M_H=450$ GeV and
$\sqrt{s}\gtrsim 3M_H$: Then the running coupling $\lambda_s^{(n)}$ is
always 
larger than $2.2$.  This choice
refers to the right plot of Fig.~\ref{sigmaoms} with $\sqrt{s}\ge 1350$ GeV.
In this region, the two-loop cross section is more than twice the size of the
one-loop cross section. The perturbativity of the result is therefore
questionable 
for values of $\lambda_s$ as low as 2.2.  Vice versa, requiring that the NLL
cross 
section and the NNLL differ by less than a factor two we find an upper bound
on a perturbative running coupling of 
\begin{equation}
\lambda_s\leq2.2\,.
\label{limit2}
\end{equation} 
The bounds from Eqs.~(\ref{limit1}) and (\ref{limit2}) give limits on the
maximal energy up to which perturbative calculations of $2\rightarrow 2$ in the
Higgs sector are possible.  These bounds are shown in Fig.~\ref{limits} where
we consider values of $\sqrt{s}$ larger than 4 TeV.

\begin{figure}[tb]
\vspace*{30pt}
\centerline{
\epsfysize=3.0in \rotate[l]{\epsffile{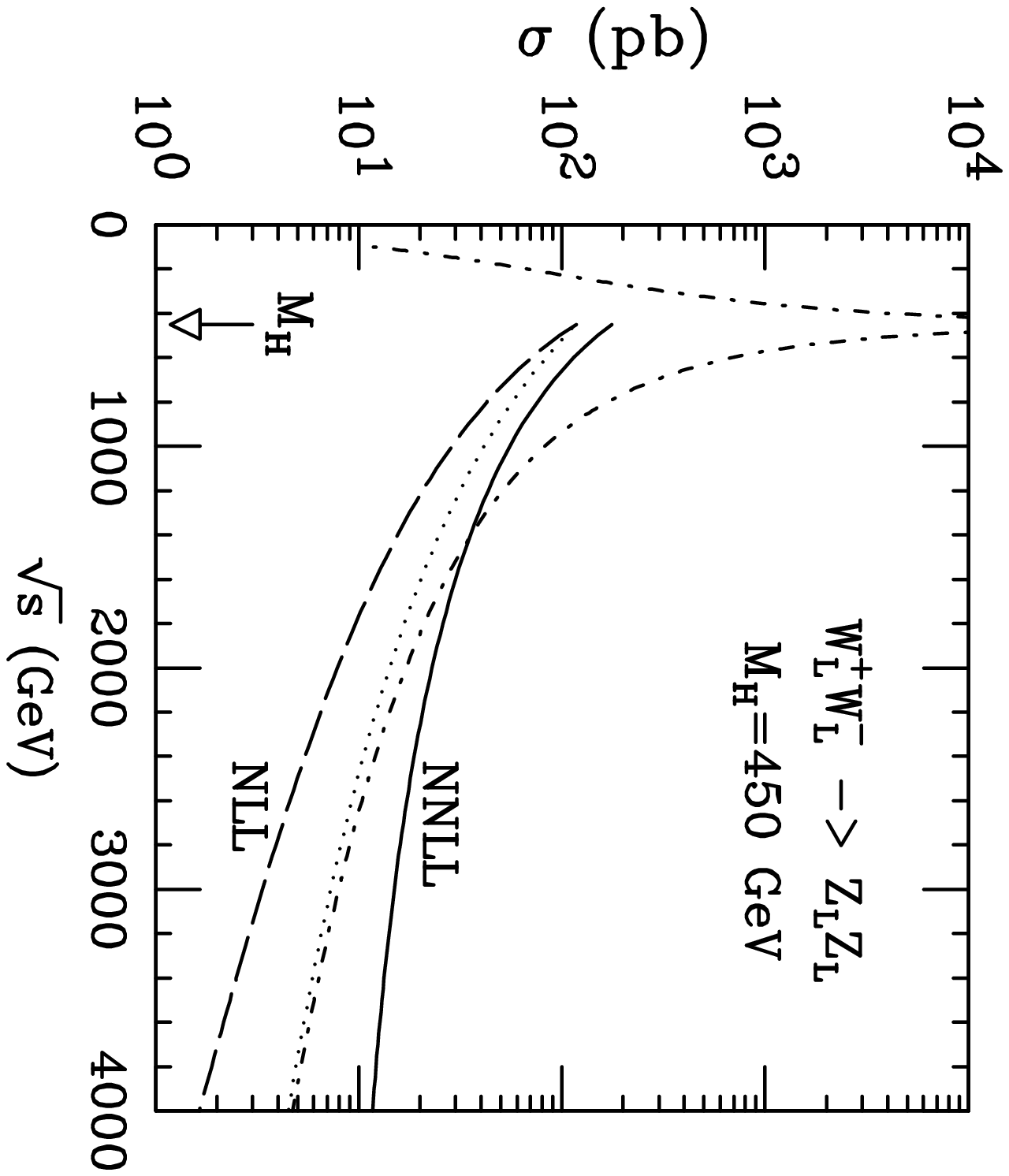}}
\hspace{.4cm}
\epsfysize=3.0in \rotate[l]{\epsffile{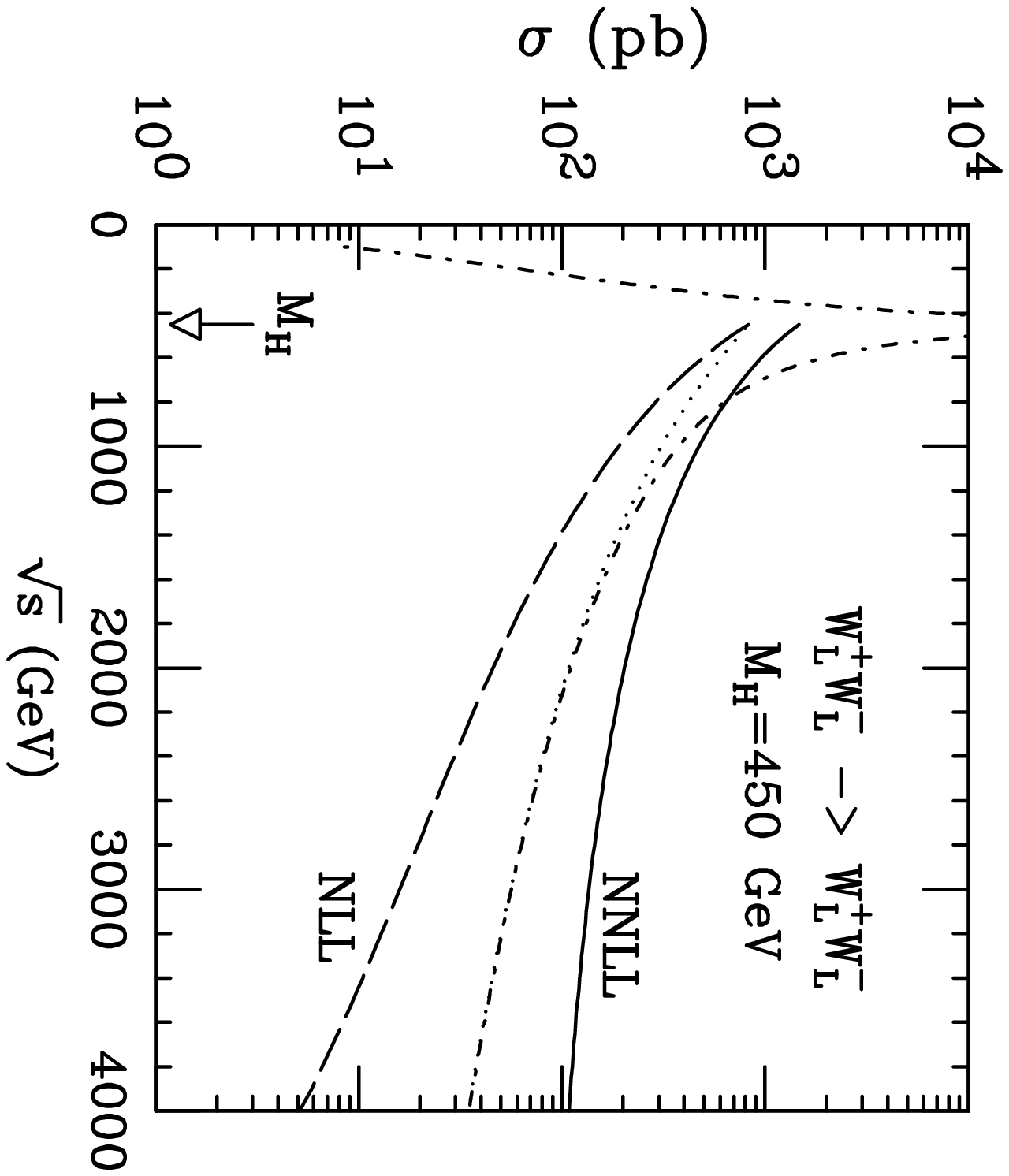}}
}
\vspace{0.15in}
\caption{
  Left plot: The cross section of the process $\protect W_L^+W_L^-\rightarrow
  Z_LZ_L$ as a function of the cms-scattering energy $\protect\sqrt{s}$ for
  $M_H=450$ GeV.  We show the NLL and NNLL curves of
  Fig.~\protect\ref{sigmaoms}, now for the whole range $M_H < \protect\sqrt{s}
  < 4000$ GeV, using a logarithmic scale for the cross section.  In addition,
  we show the exact one-loop cross section (dot-dashed curve) and the
  high-energy one-loop cross section (dotted curve), both without resummation
  of logarithms. The large corrections to the high-energy cross section change
  the exact cross section significantly, introducing large uncertainties for
  $\protect\sqrt{s}> 2M_H$.  For $\protect\sqrt{s}\protect\gtrsim 1500$ GeV
  even the NNLL perturbative result is not expected to be reliable anymore:
  perturbation theory fails.  Right plot: The cross section of the process
  $\protect W_L^+W_L^-\rightarrow W_L^+W_L^-$ as a function of the
  cms-scattering energy $\protect\sqrt{s}$ for $M_H=450$ GeV.  The curves have
  the same meaning as in the left plot, and the same conclusions apply.  }
\label{sigmaall}
\end{figure}

To illustrate the importance of our findings for future phenomenological
applications, we compare the RG results of the high-energy cross section with
the non-resummed one-loop result which contains both $\lambda$ and $\lambda v$
contributions, i.e., the exact EQT result at one loop.  In the left plot of
Fig.~\ref{sigmaall} we show the cross section of $W_L^+W_L^-\rightarrow Z_LZ_L$
for $M_H=450$ GeV.  For $\sqrt{s}<2M_H$ the cross section is dominated by the
resonant contribution from s-channel Higgs exchange. For larger values of
$\sqrt{s}$, the high-energy contribution, solely connected to the quartic
coupling, dominates the cross section; also recall Table I.  The actual size of
the cross section, however, has large uncertainties due to the bad convergence
of the RG improved cross section.

A similar behaviour can be found for all other $2\rightarrow 2$ cross sections
involving $W_L^\pm$, $Z_L$, and $H$.  Phenomenologically, the process
$W_L^+W_L^-\rightarrow W_L^+W_L^-$ is the most important one for LHC and NLC
physics. We show its cross section in the right plot of Fig.~\ref{sigmaall},
also taking $M_H=450$ GeV.  This channel has a significantly larger cross
section than the previous one.  The relative size of the radiative corrections
is slightly larger than the one of the process $W_L^+W_L^-\rightarrow
Z_LZ_L$. This is 
due to the fact that the one-loop real part of amplitude is larger in the
$W_L^+W_L^-\rightarrow W_L^+W_L^-$ channel, see Eqs.~(\ref{finwwww}) and
(\ref{finwwzz}).  We find that the actual size of the cross section has again
large 
uncertainties due to the bad convergence of the RG improved cross section if
$M_H$ is larger than ${\rm O}(450\; {\rm GeV})$ and $\sqrt{s}\gtrsim 2M_H$.
We conclude that using perturbative methods a future experimental extraction of
the quartic coupling from such cross sections is very unreliable for Higgs
masses of ${\rm O}(450 {\rm GeV})$.
 
The one-loop cross section of $W_L^+W_L^-\rightarrow W_L^+W_L^-$ was already
presented in Fig.~8 of~\cite{daw2} for $M_H=500$ GeV and $\sqrt{s}$ in the
range of 250 to 3000 GeV.  Though the authors state the importance of resumming
large logarithms using a running coupling, their cross sections are plotted
using a non-running coupling $\lambda=\lambda_{\rm OMS}$.  Our results of
Fig.~\ref{sigmaall} show that the use of the running coupling is very important
for $\sqrt{s}$ in the TeV range, and that $M_H=500$ GeV is already too large to
yield a reliable perturbative answer for $\sqrt{s}\geq 1.5$ TeV.  For
$1000\lesssim\sqrt{s}<1.5$ TeV, a NNLL perturbative answer may give a
reasonable estimate of the high-energy contribution to the cross section.


\subsection{RGE in $\overline{\rm MS}$ FORMULATION} 

Since the perturbative behaviour of the OMS high-energy amplitudes and cross
sections is not satisfactory for $M_H\gtrsim 450$ GeV, we also investigate the
RGE in the $\overline{\rm MS}$ scheme, hoping to find improved convergence.
The relevant renormalization group equation is given in Eq.~(\ref{weinberg}).
Introducing the three-loop $\overline{\rm MS}$ running coupling,
$\overline{\lambda_s}$, we can resum all $\ln(s/\mu^2)$ terms to NNLL order.
The two-loop $\ln(s/M_H^2)$ term of Eq.~(\ref{ampmsbar}) is connected to the
field anomalous dimensions.

The $\overline{\rm MS}$ three-loop running coupling differs from the OMS one
through a different value for $\lambda_s(1)$, the value of the running coupling
at the scale $\mu=\mu_0$, and a different value of the scheme dependent
three-loop coefficient $\beta_2^{\overline{\rm MS}}=12\,022.69\ldots$
\cite{vlad,beta,uli}.  In agreement with our approach in the OMS scheme, we
choose $\mu_0=M_H$.\footnote{Different choices of $\mu_0$ are discussed in
  \cite{sirzuc,uli}.} This defines $\lambda_s(1)$ as the value of the
$\overline{\rm MS}$ coupling at scale $M_H$, and it can be calculated using
Eq.~(\ref{coup}).  The resulting $\overline{\rm MS}$ value is denoted by
$\overline{\lambda_s}(1)$, and it is larger than the corresponding value in the
OMS scheme.  Consequently, the $\overline{\rm MS}$ running coupling is also
larger than the OMS running coupling.  Since the coefficients of the
$\overline{\rm MS}$ amplitudes are already in magnitude larger than the OMS
coefficients, recall Eq.~(\ref{ampmsbar}), the convergence of the
$\overline{\rm MS}$ amplitudes is worse than the convergence of the OMS
results.  To show this explicitly, we give the $\overline{\rm MS}$ result of
the two-loop high-energy cross section of $W_L^+W_L^-\rightarrow Z_LZ_L$:
\begin{eqnarray}
\sigma_{\overline{\rm MS}}^{\rm NNLL}\, 
& = & \frac{N_i^2\,N_f^2}{4\pi s}\frac{ |\vec{p_f}|}{|\vec{p_i}|}
\;\Gamma_w^8\;
\Bigl[\overline{\lambda_s}^{(3)}\Bigr]^2\,
\Biggl[\;\; 1
-\, 60\,\frac{\overline{\lambda_s}^{(3)}}{16\pi^2}\,
+\; 3370.7\,\frac{\Bigl[\overline{\lambda_s}^{(3)}\Bigr]^2}{(16\pi^2)^2}\,
+\; {\rm O}\left(\Bigl[\overline{\lambda_s}^{(3)}\Bigr]^3\right)\,\Biggr].
\label{msbarcross}
\end{eqnarray}
This result is to be compared with the OMS cross section,
Eq.~(\ref{wwzzcross}): The $\overline{\rm MS}$ corrections are significantly
larger than the OMS corrections.

Analogous to the OMS case, the one-loop perturbative cross section can become
negative if $\overline{\lambda_s}$ is too large.  In the $\overline{\rm MS}$
scheme, this happens for a running coupling of $\overline{\lambda_s}\geq
16\pi^2/60 \approx 2.6$.  Taking $M_H=450$ GeV, the one-loop $\overline{\rm 
MS}$ running coupling reaches this critical value for $\sqrt{s}=1300$ GeV,
and the two-loop $\overline{\rm MS}$ running coupling is equal to 2.6 for
$\sqrt{s}=1500$ GeV, values much lower than the OMS results of 3870 and 7800
GeV, respectively.  This indicates a large scheme dependence of the
perturbative results, another sign of the breakdown of perturbation theory.

We conclude that our OMS result -- taking $M_H= {\rm O}(450\, {\rm GeV})$ to be
the upper limit for perturbative high-energy ($\sqrt{s}\gtrsim 2M_H$)
calculations in the Higgs sector -- is strengthened by the $\overline{\rm MS}$
results.  To achieve the best apparant convergence of the perturbative results,
the OMS scheme should be employed.

Our constraints on the running coupling from a two-loop analysis can be
compared with the results 
obtained from a two-loop analysis of unitarity constraints which was carried
out in~\cite{unit}. In that paper the OMS scheme is used, and the running
coupling is defined using $\mu_0\approx 0.7 M_H$ as suggested by~\cite{sirzuc}.
This different choice of $\mu_0$ corresponds to a resummation of constant
terms, resulting in larger magnitudes of the subleading coefficients of the
perturbative amplitudes compared to our OMS result with $\mu_0=M_H$.  In fact,
the bounds received in~\cite{unit} are similar to the $\overline{\rm MS}$
constraints found here, i.e., somewhat more stringent than the OMS results of
the previous subsection.

\section{Summary}

The Higgs quartic coupling dominates the cross sections of elastic
$2\rightarrow 2$ processes involving longitudinally polarized gauge bosons and
the Higgs boson for $\sqrt{s}\gtrsim 1.5$ -- $2M_H$. Using perturbative
amplitudes up to two loops and considering cms energies of up to 4 TeV, we find
the cross sections to have large 
uncertainties if $\sqrt{s}\gtrsim 2M_H$ and $M_H\gtrsim 450$ GeV.  This is due
to a unsatisfactory convergence of the perturbative series in the OMS scheme.
The breakdown of perturbation theory is due to large logarithmic terms (which
lead to a large running coupling) as well as large
non-logarithmic contributions to the perturbative cross section. One-loop and
two-loop high-energy cross sections 
differ factors of two or more if $M_H$ is ${\rm O}(450\, {\rm
GeV})$. Simultaneously, the cross section exhibits a 
large renormalization scheme dependence as seen in the comparison of OMS and
$\overline{\rm MS}$ results, with the OMS scheme giving a better convergence of
the perturbative series. The large uncertainties in the perturbative cross
sections will 
inhibit the analysis of the quartic Higgs coupling in the case of a heavy
Higgs, $M_H\gtrsim 450$ GeV, decreasing the experimental sensitivity to physics
beyond the Standard Model significantly.

If $M_H\lesssim 350$ GeV, the apparant convergence of the perturbation series
is satisfactory, still assuming $\sqrt{s}\lesssim 4$ TeV.  However, for 
energies $\sqrt{s}> 2M_H$ it is essential to resum the leading logarithmic
energy dependence to all orders using the running coupling.  NLL terms are
subdominant, and the Standard Model cross sections for longitudinal gauge boson
scattering can be predicted with errors of less than 10\% if the NNLL
approximation is used,  sufficient for LHC and future NLC
experiments.

\section{ACKNOWLEDGMENTS}
The author would like to thank L.~Durand and U.~Nierste for many useful
discussions.  Further stimulation and information was provided through
discussions with A.~Buras, R.~Hempfling, B.~Kniehl, M.~Lindner, and A.~Sirlin.
This work was partially supported by the Deutsche Forschungsgemeinschaft (DFG)
under contract number Li519/2-1.
\appendix

\section{Angular dependent result of $W_L^+W_L^-\rightarrow Z_LZ_L$ }
\label{angular}

We write the high-energy transition amplitudes of the process
$W_L^+W_L^-\rightarrow Z_LZ_L$ in the general form 
\begin{eqnarray}
\mbox{\boldmath${\cal M}$}
&=& \,{ c_{10}}\,{ \lambda} +  \biggl[ \! \,{ c_{21}}\,
{\rm ln} \left( \! \,s/\mu^2\,
 \!  \right)  + { c_{20}} + { d_{20}}\,\Bigl[\,{\rm ln}(\,{-t/s}\,) + 
{\rm ln}(\,{-u/s}\,)\,\Bigr]\, \!  \biggr] \,\frac{ \lambda^2 }{16\pi^2}
\nonumber\\
& &+  \left[ {\vrule height0.93em width0em depth0.93em} \right. \! \! 
\;\;  { c_{32}}\,{\rm ln}^2 \left( \! \,s/\mu^2\, \!  \right)  
+ { c_{31}}\,{\rm ln} \left( \! \,
s/\mu^2\, \!  \right)  
+ { c'_{31}}\,{\rm ln} \left( \! \,
s/M_H^2 \!  \right) \;
+ { c_{30}}\nonumber\\
& &\;\;\;\;+ \left[{ d_{31}}\,{\rm ln} \left( \! \,s/\mu^2\, \!  \right) 
+ { d_{30}}\,\right]\,\Bigl[\,{\rm ln}(\,{-t/s}\,) + {\rm
ln}(\,{-u/s} \,)\,\Bigr] \nonumber\\
 & & \;\;\;\;+\; { e_{30}} \left[ \; \,{\rm ln}^2(\,{-t/s}\,)+ {\rm ln}^2(
\,{-u/s}\,)\, \!  \right]  
 \! \! \left. {\vrule height0.93em width0em depth0.93em} \right]
 \frac{\lambda^3}{(16\pi^2)^2},
\label{genform}
\end{eqnarray}
a form which is adequate for all $t\leftrightarrow u$ symmetric channels.  No
renormalization scheme has been specified.  In the OMS-renormalization scheme,
the scale $\mu$ is defined as $\mu=M_H$.

The quantities $s$, $-t$, and $-u$ are real and positive in the physical
region. The terms $\ln(-t/s)$ and $\ln(-u/s)$ are a function of only the
scattering angle. In the center-of-mass system we have
$-t/s=(1+\cos{\theta_{cm}})/2$, $-u/s=(1-\cos{\theta_{cm}})/2$.

The coefficients $c_{nm}$ correspond to terms independent of the scattering
angle. Coefficients $d_{nm}$ and $e_{nm}$ refer to terms containing an angular
dependence as $\ln(1\pm\cos\theta)$ and $\ln^2(1\pm\cos\theta)$, respectively.
The index $n$ refers to the order in perturbation theory, $\lambda^n$. The
index $m$ indicates the power in $\ln(s)$.  In the OMS scheme, the coefficients
can be calculated using the results of~\cite{mah}. The $\overline{\rm MS}$
coefficients are then derived as outlined in the text.  To indicate the scheme
dependence of the different coefficients as well as the connection of certain
coefficients to the beta function, we give the explicit analytical result in
the case of $W_L^+W_L^-\rightarrow Z_LZ_L$.  Coefficients with a bar indicate
$\overline{\rm MS}$ quantities, others are OMS quantities. Recall that
$\overline{\beta_0}=\beta_0$ and $\overline{\beta_1}=\beta_1$.
\begin{eqnarray}
{\rm Tree\: level:\:}&&\nonumber\\
c_{10} & \, = \, & \overline{c}_{10}\, =\, -2\, .  \\
{\rm One\:\: loop:\:\:}&&\nonumber\\
c_{21} & \, = \, & \overline{c}_{21}\, 
= \,\frac{\beta_0}{2}\,c_{10}\,= \, -24\, ,   \\
c_{20} & \, = \, & 2+6\pi\sqrt{3} + 16\pi \, i \,\approx\, 34.648 + 50.265 \, i
\, , \\ 
\overline{c}_{20}& \, = \, & 52 + 16\pi \, i  \\
d_{20} & \, = \, & \overline{d}_{20}\, = \, -4 \, . \\
{\rm Two\:\: loops:}&&\nonumber\\
c_{32} & \, = \, & \overline{c}_{32}\, = \,
\left(\frac{\beta_0}{2}\right)^2c_{10}\,=\, -288 \, ,  \\
c_{31} & \, = \, & \frac{\beta_1}{2}c_{10} + \beta_0\,c_{20}
\, = \, 360+144\pi\sqrt{3} + 384\pi \, i 
\,\approx\, 1143.561 + 1206.372 \, i  \, , \\ 
\overline{c}_{31}& \, = \, &  
\frac{\beta_1}{2}\overline{c}_{10} + \beta_0\,\overline{c}_{20}
\, = \, 1560 + 384\pi \, i \, ,\\
c'_{31} & \, = \, & \overline{c'}_{31}\, = \, -24 \, ,  \\
d_{31} & \, = \, & \overline{d}_{31}\, = \, \beta_0\,d_{20}
\, = \,  -96 \, ,  \\
c_{30} & \, = \, & -74-224\zeta(2)+ 180\zeta(3) - 384{\bf Cl}\sqrt{3}
             +138\pi\sqrt{3} -96\pi{\bf Cl} - 324K_5 \nonumber\\
  &&\; - (176\pi + 96\pi^2\sqrt{3}) \, i \,\approx\, -755.583 - 2194.007\, i 
\, ,\\
\overline{c}_{30}& \, = \, & 
             -2524 + 824\zeta(2) + 48{\bf Cl}\sqrt{3} - 976\pi \, i
             \,\approx\, - 1084.194 - 3066.194 \, , \\
d_{30} & \, = \, & 80 + 24\pi\sqrt{3} \,\approx\, 210.594 \, , \\
\overline{d}_{30}& \, = \, & 280 \, ,  \\
e_{30} & \, = \, & \overline{e}_{30}\, = \, -48\, ,   
\end{eqnarray}
where the constants $\zeta(2)\,,\zeta(3)\,,{\bf Cl}$, and $K_5$ 
are defined following Eq.~(\ref{coup}) in the text.

\section{Partial-wave projected amplitudes}
\label{resj0}

Here we give the OMS high-energy results for the $j=0$ partial-wave projected
amplitudes for all $2\rightarrow2$ channels considered. They are derived using
the results of~\cite{mah}. The running coupling is understood to be the
two-loop running coupling, $\lambda_s\equiv\lambda_s^{(2)}$, defined in
Eqs.~(\ref{pert2lp}), (\ref{ii11}).  If the three-loop running coupling becomes
available, it can be used without a change in the coefficients presented here,
leading to a complete resummation of all NNLL terms. The overall factor $\left(
  {4|\vec{p_i}||\vec{p_f}|}/{s} \right)^{1/2}$ is taken to be unity
(high-energy limit).

\begin{eqnarray}
&&{W_L^+W_L^-\rightarrow W_L^+W_L^-:} \rule{0pt}{20pt}\nonumber\\
&&\;{{\bf a}_{0}(s)}=\;\;\frac{ -8\,{ \lambda_s^{(2)}}
\Gamma_w^4}
{32\pi} \rule{0pt}{28pt}\;\;
\Biggl[\,1 -\,  \Bigl(\,  24.32 + 15.71\,i\, \Bigr) 
                \,\frac{ \lambda_s^{(2)}}{16\pi^2}
           +\,  \Bigl(\, 1233.6 + 713.9\,i\, \Bigr)
                \biggl[\frac{\lambda_s^{(2)}}{16\pi^2}\biggr]^2\;
\Biggr]. 
\label{finwwww}\\
&&{Z_LZ_L\rightarrow Z_LZ_L: }\rule{0pt}{20pt}\nonumber\\
&&\;{{\bf a}_{0}(s)=\;\frac{ -12\,{ \lambda_s^{(2)}}
\Gamma_w^4}
{64\pi} \rule{0pt}{28pt}\;
\Biggl[\,1 -\,  \Bigl(\,  25.32 + 12.57\,i\, \Bigr) 
                \,\frac{ \lambda_s^{(2)}}{16\pi^2}
           +\,  \Bigl(\, 1416.7 + 586.2\,i\, \Bigr)
                \biggl[\frac{\lambda_s^{(2)}}{16\pi^2}\biggr]^2\;
\Biggr]. }\\
&&W_L^+W_L^-\rightarrow Z_LZ_L: \rule{0pt}{20pt}\nonumber\\
&&\;{{\bf a}_{0}(s)=\;\;\frac{ -4\,{ \lambda_s^{(2)}}
\Gamma_w^4}
{32\pi\sqrt{2}} \rule{0pt}{28pt}\;\;
\Biggl[\,1 -\,  \Bigl(\,  21.32 + 25.13\,i\, \Bigr) 
                \,\frac{ \lambda_s^{(2)}}{16\pi^2}
           +\,  \Bigl(\, 684.3 + 1097.0\,i\, \Bigr)
                 \biggl[\frac{\lambda_s^{(2)}}{16\pi^2}\biggr]^2\;
\Biggr]. }
\label{finwwzz}\\
&&HH\rightarrow W_L^+W_L^-: \rule{0pt}{20pt}\nonumber\\
&&\;{{\bf a}_{0}(s)}=\frac{ -4\,{ \lambda_s^{(2)}}
\Gamma_w^2\Gamma_H^2}
{32\pi\sqrt{2}} \rule{0pt}{28pt}
\Biggl[\,1 -\,  \Bigl(\,  19.21 + 25.13\,i\, \Bigr) 
                \,\frac{ \lambda_s^{(2)}}{16\pi^2}
           +\,  \Bigl(\,  606.5 + 1043.7\,i\, \Bigr)
                \biggl[\frac{\lambda_s^{(2)}}{16\pi^2}\biggr]^2\;
\Biggr]. \\
&&HH\rightarrow Z_LZ_L: \rule{0pt}{20pt}\nonumber\\
&&\;{{\bf a}_{0}(s)}=\frac{ -4\,{ \lambda_s^{(2)}}
\Gamma_w^2\Gamma_H^2}
{64\pi} \rule{0pt}{28pt}
\Biggl[\,1 -\,  \Bigl(\,  19.21 + 25.13\,i\, \Bigr) 
                \,\frac{ \lambda_s^{(2)}}{16\pi^2}
           +\,  \Bigl(\,  606.5 + 1043.7\,i\, \Bigr)
                \biggl[\frac{\lambda_s^{(2)}}{16\pi^2}\biggr]^2\;
\Biggr]. \\
&&HH\rightarrow HH:\rule{0pt}{20pt}\nonumber\\ 
&&\;{{\bf a}_{0}(s)}=\;\;\frac{ -12\,{ \lambda_s^{(2)}}
\Gamma_H^4}
{64\pi} \rule{0pt}{28pt}\;
\Biggl[\,1 -\,  \Bigl(\,  21.09 + 12.57\,i\, \Bigr) 
                \,\frac{ \lambda_s^{(2)}}{16\pi^2}
           +\,  \Bigl(\, 1248.6 + 532.9\,i\, \Bigr)
                \biggl[\frac{\lambda_s^{(2)}}{16\pi^2}\biggr]^2\;
\Biggr]. \\
&&H Z_L\rightarrow HZ_L: \rule{0pt}{20pt}\nonumber\\
&&\;{{\bf a}_{0}(s)}=\frac{ -4\,{ \lambda_s^{(2)}}
\Gamma_w^2\Gamma_H^2}
{32\pi} \rule{0pt}{28pt}\:
\Biggl[\,1 -\,  \Bigl(\,  25.21 \,+\; 6.28\,i\, \Bigr) 
                \,\frac{ \lambda_s^{(2)}}{16\pi^2}
           +\,  \Bigl(\, 1692.3 + 317.5\,i\, \Bigr)
                \biggl[\frac{\lambda_s^{(2)}}{16\pi^2}\biggr]^2\;
\Biggr]. 
\end{eqnarray}
The explicit $\ln(s)$ dependence up to two loops can be obtained by expanding
$\lambda_s^{(2)}$, $\Gamma_w$, and $\Gamma_H$ in power of
$\lambda_s(1)\equiv\lambda_{\rm OMS}$, dropping terms of ${\rm O}(\lambda_{\rm
OMS}^4)$. The constant terms given above will not receive any corrections since
our definition of the running coupling, Eq.~(\ref{pert2lp}), does not involve
any resummation of constant terms.

The $j=0$ partial-wave projected amplitude can be used to calculate the total
cross section in very good approximation.  Since the tree-level high-energy
amplitude is independent of the scattering angle, only the $j=0$ partial wave
is non-zero.  Including higher order corrections, the matrix elements depend on
the scattering angle, but with rather small coefficients.  The $j=0$
partial-wave projected cross section remains dominant and is an excellent
approximation of the total two-loop cross section:
\begin{eqnarray}
\sigma\, 
 &\approx& \;\frac{4\pi}{|\vec{p_i}|^2} |{\bf a}_{j=0}|^2 \, ,
\label{sigapprox}
\end{eqnarray}
where $|{\bf a}_{j=0}|^2$ is understood to be expanded in powers of
$\lambda_s$, dropping terms of ${\rm O}\left([\lambda_s^{(2)}]^5\right)$.  At
one loop, neglecting terms of ${\rm O}(\lambda_s^4)$, the total cross section
is actually {\it equal} to the $j=0$ cross section. At two loops, the
$\lambda_s^4$ coefficients are not identical anymore, but their numerical
difference is much less than 1\%.  The $j=0$ partial-wave projected cross
section of the channel $W_L^+W_L^-\rightarrow Z_LZ_L$, for example, has a
two-loop coefficient of $2\,455.1\,$, whereas the total cross section, given
in Eq.~(\ref{wwzzcross}), has the coefficient $2\,457.9\,$.



\begin{table}
\squeezetable
\begin{tabular}
{|@{$\;\;\;\;$}c@{$\;\;\;\;$}|
 @{$\;\;\;\;$}l@{$\;\;\;$}|
 @{$\;\;\;$}d|
 @{$\;\;\;$}d|
 @{$\;\;\;$}d|} 
$M_H$ (GeV) &  
\hspace{1mm}{$\sqrt{s}\;$} (GeV) & 
{$\Delta (a_{j=0}^{\rm tree})\;\;$} &
{$\Delta ({\rm Re}\, a_{j=0}^{\rm 1lp})\;\;$}  & 
{$\Delta ({\rm Im}\, a_{j=0}^{\rm 1lp})\;\;$}  \\ 
\hline 
  200  &  \hspace{0.7mm}  300  &    44.4  &    46.4  &    39.2 \\ 
  200  &  \hspace{0.7mm}  400 $\;\;(2M_H)$ &    25.0  &    27.1  &     6.1 \\ 
  200  &  \hspace{0.7mm}   500  &    16.0  &    17.8  &    -6.5 \\ 
  200  &  \hspace{0.7mm}  600 $\;\;(3M_H)$ &    11.1  &    12.7  &   -11.2 \\ 
  200  &  \hspace{0.7mm}  800 $\;\;(4M_H)$ &     6.3  &     7.4  &   -12.6 \\ 
  200  &   1000 $\;\;(5M_H)$ &     4.0  &     4.9  &   -11.1 \\ 
  200  &   2000  &     1.0  &     1.4  &    -4.7 \\ 
  200  &   3000  &     0.4  &     0.7  &    -2.5 \\ 
  200  &   4000  &     0.2  &     0.4  &    -1.6 \\ 
\hline 
  500  &   \hspace{0.7mm}  750  &    44.4  &    56.2  &    39.2 \\ 
  500  &   1000 $\;\;(2M_H)$ &    25.0  &    36.9  &     6.1 \\ 
  500  &   1250  &    16.0  &    26.3  &    -6.5 \\
  500  &   1500 $\;\;(3M_H)$ &    11.1  &    19.8  &   -11.2 \\ 
  500  &   2000 $\;\;(4M_H)$ &     6.3  &    12.4  &   -12.6 \\ 
  500  &   2500 $\;\;(5M_H)$ &     4.0  &     8.6  &   -11.1 \\
  500  &   3000  &     2.8  &     6.3  &    -9.3 \\ 
  500  &   4000  &     1.6  &     4.0  &    -6.5 \\ 
\hline 
  800  &   1200  &    44.4  &    71.7  &    39.2 \\ 
  800  &   1600 $\;\;(2M_H)$ &    25.0  &    51.2  &     6.1 \\ 
  800  &   2000  &    16.0  &    38.0  &    -6.5 \\ 
  800  &   2400 $\;\;(3M_H)$ &    11.1  &    29.1  &   -11.2 \\ 
  800  &   2500  &    10.2  &    27.4  &   -11.8 \\ 
  800  &   3000  &     7.1  &    20.5  &   -12.7 \\ 
  800  &   3200 $\;\;(4M_H)$  &     6.3  &    18.5  &   -12.6 \\ 
  800  &   4000 $\;\;(5M_H)$ &     4.0  &    12.7  &   -11.1 \\ 
  \end{tabular} 
\vspace{4mm}
\caption{The relative difference between the exact EQT amplitude and the
high-energy EQT amplitude, $\Delta(a_{j=0})=(a_{j=0}^{\rm exact}-a_{j=0}^{\rm
high})/a_{j=0}^{\rm 
exact}$, in percent. The first two columns give the values for the Higgs mass
and the cms energy in units of GeV. The third column shows the difference at
tree level, the forth and fifth column the one-loop difference separated into
real and imaginary part, respectively.}
\label{table1}
\end{table}

\newpage

\newpage

\newpage

\newpage

\newpage

\newpage

\newpage

\end{document}